\documentclass[twocolumn,amsmath, amssymb, aps, pra, superscriptaddress]{revtex4-1}

\usepackage{amsmath,amssymb,amsfonts,bm}
\usepackage{graphicx,epstopdf}  
\usepackage{color}
\usepackage{xcolor}
\usepackage{extarrows}
\usepackage{enumitem}
\usepackage{empheq}
\usepackage{hyperref}

\newcommand{\la}{\langle}
\newcommand{\ra}{\rangle}

\newcommand{\Pade}{\text{Padé}\ }

\newcommand{\Ht}{H_{_{\rm T}}}
\newcommand{\Hs}{H_{_{\rm S}}}
\newcommand{\He}{H_{_{\rm E}}}
\newcommand{\Hse}{H_{_{\rm SE}}}

\newcommand{\bst}{\bm{s}_t}
\newcommand{\btheta}{\bm{\theta}}

\newcommand{\be}{\begin{equation}}
\newcommand{\ee}{\end{equation}}
\newcommand{\bsube}{\begin{subequations}}
\newcommand{\esube}{\end{subequations}}
\newcommand{\Eq}[1]{Eq.\,(\ref{#1})}

\newcommand{\RN}[1]{%
  \textup{\uppercase\expandafter{\romannumeral#1}}%
}

\begin{document}

\title{Simulating non-Markovian open quantum dynamics by exploiting physics-informed neural network}

\author{Long Cao} 
\affiliation{Hefei National Research Center for Physical Sciences at the Microscale, 
University of Science and Technology of China, Hefei, Anhui 230026, China}

\author{Liwei Ge} 
\affiliation{Hefei National Research Center for Physical Sciences at the Microscale, 
University of Science and Technology of China, Hefei, Anhui 230026, China}

\author{Daochi Zhang}
\affiliation{State Key Laboratory of Porous Materials for Separation and Conversion \& MOE Key Laboratory of Computational Physical Sciences \& Department of Chemistry, Fudan University, Shanghai 200438, China}

\author{Yao Wang}
\affiliation{Hefei National Research Center for Physical Sciences at the Microscale, 
University of Science and Technology of China, Hefei, Anhui 230026, China}

\author{Rui-Xue Xu}
\affiliation{Hefei National Research Center for Physical Sciences at the Microscale, 
University of Science and Technology of China, Hefei, Anhui 230026, China}
\affiliation{Hefei National Laboratory, Hefei, Anhui 230088, China}

\author{YiJing Yan}
\affiliation{Hefei National Research Center for Physical Sciences at the Microscale, 
University of Science and Technology of China, Hefei, Anhui 230026, China}

\author{Xiao Zheng} \email{xzheng@fudan.edu.cn}
\affiliation{State Key Laboratory of Porous Materials for Separation and Conversion \& MOE Key Laboratory of Computational Physical Sciences \& Department of Chemistry, Fudan University, Shanghai 200438, China}
\affiliation{Hefei National Laboratory, Hefei, Anhui 230088, China}

\date{Submitted on March~8, 2026; Resubmitted on April~26, 2026}

\begin{abstract}

This work integrates the physics-informed neural network (PINN) approach into the neural quantum state framework to simulate open quantum system dynamics, to circumvent the computationally expensive time-dependent variational principle required in conventional variational methods. The proposed PINN-DQME method employs time-encoded neural networks within a time-domain decomposition strategy to represent the evolution governed by the dissipaton-embedded quantum master equation (DQME). We implement and validate this approach in the single-impurity Anderson model, benchmarking the PINN-DQME results against the numerically exact hierarchical equations of motion. The PINN-DQME method demonstrates high accuracy in simulating quantum dissipative dynamics at high temperatures, where non-Markovian effects are weak. However, for strongly non-Markovian dynamics at low temperatures, it encounters challenges with error accumulation during time propagation, highlighting an area for future refinement in applying PINNs to complex quantum dynamical settings.

\end{abstract}

\maketitle

\section{Introduction} \label{sec:intro}

Many-body open quantum systems (OQSs) have attracted wide attention due to their profound applications in diversified fields, including physics, chemistry, materials science, and life sciences. These applications cover various fields such as coherent energy transfer in biological photosystems \cite{Engel2007,doi:10.1021/jz900062f,doi:10.1126/science.1235820}, 
charge transfer in molecular aggregates \cite{6395f2ce-ffb5-3279-8695-df26ad69dc2d,doi:10.1021/acs.chemrev.7b00581}, 
electron transport in single molecular junctions \cite{10.1063/1.5003306,Uzma2021}, 
multidimensional coherent spectroscopy of condensed phase materials \cite{10.1063/1.4994987,Bruder2018}, 
correlated quantum matter for quantum information and computation
\cite{Monroe2002,head2020quantum},
and precise measurement and control of local spin states in surface-adsorbed molecules \cite{RevModPhys.94.045008}.
Accurate characterization of environmental dissipations and system correlations is crucial for the studies of OQSs \cite{wilson1975renormalization,doi:10.1126/science.280.5363.567,doi:10.1126/science.1113449,li2020molecular,PhysRevLett.111.086601,doi:10.1126/science.aay6779,doi:10.1126/science.abg8223,Ding25084114}.

Numerous approaches have been developed to simulate the non-Markovian dynamics of many-body OQSs \cite{Tam18030402,Lam193721,PhysRevLett.82.1801,10.1063/1.1647528,PhysRevLett.123.050601,PhysRevLett.88.170407,suess2014hierarchy,PhysRevLett.115.266802,PhysRevLett.130.186301,PhysRevB.87.115115,PhysRevLett.88.256403,RevModPhys.92.011001,10.1063/1.3173823,RevModPhys.93.015008,li2023dissipatons,Yan14054105,li2024toward,moix2013hybrid,duan2017zero,hsieh2018unified,10.1063/1.1580111}. Among these, the dissipaton-embedded quantum master equation (DQME)  \cite{Yan14054105,li2024toward} has emerged as a promising method, offering a compact representation of quantum dissipative dynamics and excellent transferability. Its second-quantized formulation has been demonstrated to be equivalent to the extensively employed hierarchical equations of motion (HEOM) \cite{Tan89101,YAN2004216,Jin08234703,PhysRevLett.109.266403}, and the dissipaton picture shares common features with the pseudomode theory \cite{PhysRevA.55.2290,Tam18030402,Lam193721,cirio2023pseudofermion,Lin251289,PhysRevResearch.2.043058}.
However, these methods all confront an inherent exponential wall: their computational costs scale exponentially with both the system size and the complexity of environmental memory.


To overcome this problem, one common approach is to employ a variational ansatz to represent the dynamical variables in a quantum dynamical equation, such as the reduced density operator (RDO) and auxiliary density operators (ADOs) in the HEOM method and the reduced density tensor (RDT) in the DQME method.
Tensor network states (TNSs) \cite{verstraete2008matrix,schollwock2011density,orus2014practical,shi2018efficient} and neural quantum states (NQSs) \cite{doi:10.1126/science.aag2302,PhysRevLett.125.100503,PhysRevLett.122.250502,PhysRevLett.122.250501,PhysRevLett.122.250503,PhysRevB.99.214306,cao2026simulating,ye2025simulating} approaches have been developed to efficiently represent these time-dependent dynamical variables.
For instance, matrix product states (MPSs) have been employed to compress the ADOs in HEOM \cite{shi2018efficient,ke2022hierarchical,ke2023tree,preston2025nonadiabatic}, while restricted Boltzmann mechines (RBMs) have been used to represent the RDT in DQME \cite{cao2026simulating}.
The resulting MPS-HEOM and RBM-DQME approaches have achieved significant  reduction in the number of required dynamical variables. 
Recently, operator learning techniques have also been explored, offering an alternative approach for simulating OQS dynamics by utilizing machine learning \cite{zhang2024artificial,zhang2025neural}.


When simulating the dynamics of OQSs with variational methods, the time-dependent variational principle (TDVP) \cite{doi:10.1126/science.aag2302,PhysRevLett.125.100503,PhysRevLett.122.250502,reh2021time} is employed to project the reduced system dynamics onto the parameter space of a chosen variational ansatz. This projection yields equations of motion for the variational parameters. A key computational burden in this approach is the necessity to compute the time derivatives of these variational parameters at each incremental time step, which significantly hinders the broader development and application of variational methods \cite{cao2026simulating}. For instance, applying TDVP within the NQS framework requires solving a large system of linear equations at each time step. Both the construction of the coefficient matrix and the subsequent solution of the system of linear equations are time-consuming \cite{cao2026simulating} due to the high Monte Carlo sampling number for high-accuracy requirements and the iterative optimizer if the linear problem is large.


An alternative method is provided by physics-informed neural networks (PINNs), which use neural networks to represent solutions of  partial differential equations (PDEs)
by reformulating the governing equations as a loss function optimization problem. This is achieved by embedding the mathematical structure of the PDE (e.g., through differential operators such as $\partial_t, \partial_x$) directly into the neural network, and defining a loss that incorporates residuals from the governing equations and boundary conditions. For time-dependent PDEs, time $t$ is treated as an additional input node, allowing a single network to represent the solution over the entire temporal domain. 
Since the original PINN formulation \cite{raissi2018deep}, numerous variants have been developed, including physics-constrained neural networks \cite{liu2021dual}, hp-variational PINNs \cite{kharazmi2021hp}, and conservative PINNs \cite{jagtap2020conservative}. 
PINNs have found broad application across fields such as fluid dynamics, optics, electromagnetism, and geoscience \cite{cuomo2022scientific}. However, their use in quantum mechanics remains largely limited to non-dissipative situations, such as solving the Schr\"{o}dinger equation for the transverse field Ising model, \cite{sinibaldi2026time,van2025many} or addressing data-driven problems \cite{ullah2024physics,ullah2025machine,atif2026toward}.  While recent attempts have been made to apply PINNs to quantum master equations for OQSs \cite{dugan2023q}, the results have shown rather limited success. Nevertheless, the PINN approach offers a potential computational advantage: by encoding temporal evolution directly into the network inputs, it avoids the need to update network parameters at each time step, which is required for NQS approaches that rely on the TDVP. This feature could circumvent the key computational bottlenecks associated with NQS methods.



In this work, we propose a PINN-based framework that represents the dynamics of OQSs governed by DQME via time-independent neural networks, aiming at avoiding the time-consuming TDVP in conventional variational methods.  To the best of our knowledge, the study represents the first demonstration of applying PINNs to the dissipative dynamics of many-body OQSs, opening a promising avenue for efficient and scalable quantum dynamics simulation.

The remainder of this paper is organized as follows. In Section~\ref{subsec:dqme}, we provide the details of the DQME method. In Section~\ref{subsec:pinn}, we present the neural network architecture and the loss function utilized to optimize the neural network to characterize the dynamics of OQSs. Then in Section~\ref{sec:result}, we discuss the results of our method. Finally, Section~\ref{sec:conclude} summarizes this work.

\section{Methodology} \label{sec:method}
\subsection{Fermionic DQME formalism}  \label{subsec:dqme}

We consider an Anderson impurity model \cite{And6141} in which the impurity is linearly coupled to the environment consisting of several noninteracting electron reservoirs. 
The Hamiltonian of the total system is described by (with $e = \hbar  = 1$)
\begin{align}
\Ht & = \Hs + \He + \Hse, \\
\He & = \sum_\alpha \sum_{k,s}  \epsilon_{\alpha ks} \hat{d}^{\dagger}_{\alpha ks}\hat{d}_{\alpha ks}, \\
\Hse & = \sum_{\nu=1}^{N_{_{\rm S}}} \sum_{\alpha, s}
	\left(\hat{c}_{\nu s}^{\dagger}\hat{F}_{\alpha \nu s} + \hat{F}_{\alpha \nu s}^{\dagger}\hat{c}_{\nu s}\right), 
\end{align}
where $\Hs$ represents the system of primary interest, $\He$ denotes the environment, and $\Hse$ describes the system-environment coupling. 
In these terms, $\hat{c}^\dag_{\nu s}$ ($\hat{c}_{\nu s}$) denotes the creation (annihilation) operator for an electron of spin $s$ at the $\nu$-th energy level of the system,  $\hat{d}^\dag_{\alpha k s}$ ($\hat{d}_{\alpha k s}$) denotes the creation (annihilation) operator for an electron of spin $s$ at the $k$-th state of the $\alpha$-th reservoir. 
%
$\hat{F}_{\alpha \nu s}=\sum_k t_{\nu \alpha k s}\hat{d}_{\alpha k s}$, where $t_{\nu \alpha k}$ denotes the coupling strength between the system's $\nu$-th state and the $\alpha$-reservoir's $k$-th state for the electrons with spin $s$.
%

Since a reservoir of  noninteracting fermionic particles linearly coupled to the system follows Gaussian statistics, its influence on the reduced system dynamics is entirely captured by the reservoir hybridization correlation functions \cite{Jin08234703}, which are defined as
\begin{align}
    C_{\alpha \nu s}^{+}(t-\tau) & \equiv \langle 
        \hat{F}_{\alpha \nu s}^{\dagger}(t) \hat{F}_{\alpha \nu s}(\tau) \rangle_{_{\rm E}}, \nonumber \\ 
    C_{\alpha \nu s}^{-}(t-\tau) & \equiv \langle
        \hat{F}_{\alpha \nu s}(t) \hat{F}^{\dagger}_{\alpha \nu s}(\tau) \rangle_{_{\rm E}},
\end{align}
where $\hat{F}_{\alpha \nu s}(t) = e^{i \He t}\hat{F}_{\alpha \nu s}e^{-i \He t}$, and $\la \cdot \ra_{_{\rm E}} = {\rm tr}( \cdot \rho^{\rm eq}_{_{\rm E}})$ with $\rho^{\rm eq}_{_{\rm E}}$ being the density matrix of the decoupled environment in thermal equilibrium. 
The reservoir correlation functions are related to the hybridization spectral functions through the fluctuation-dissipation theorem:
\begin{equation}
    C_{\alpha \nu s}^{\sigma}(t) = 
    \int_{-\infty}^{\infty} {\rm d} \omega \, e^{\sigma i \omega t} 
    f^{\sigma}_{\alpha}(\omega) J_{\alpha \nu s}(\omega), \label{eqn:FD}
\end{equation}
where $f^{\sigma}_{\alpha}(\omega) = 1/(1+e^{\sigma \beta_{\alpha} (\omega-\mu_{\alpha})})$, and $\beta_{\alpha} = 1/(k_{\rm B}T)$ is the inverse temperature of the $\alpha$-reservoir. 
The hybridization spectral functions of the $\alpha$-reservoir are defined as
\begin{equation}
    J_{\alpha \nu s}(\omega) \equiv \pi \sum_k \left\lvert t_{\nu \alpha k s} \right\rvert^2
    \delta(\omega-\epsilon_{\alpha k s}).
\end{equation}
In this work, we adopt a Lorentzian form for the hybridization spectral functions, i.e.,
\begin{equation}
    J_{\alpha \nu s}(\omega) = \frac{\Gamma_{\nu s\alpha} W_\alpha^2 }{(\omega-\Omega_{\alpha})^2+W_\alpha^2}.  
\end{equation}
Here, $\Gamma_{\nu s \alpha}$ is the hybridization strength between the system's $\nu$-th state and the $\alpha$-reservoir for the electrons with spin $s$, $\Gamma_{\nu s} = \sum_\alpha \Gamma_{\nu s \alpha}$, and $\Omega_\alpha$ and $W_\alpha$ are the band center and width of the $\alpha$-reservoir, respectively. 
For simplicity, we take $\Omega_\alpha = \mu_\alpha$, where $\mu_\alpha$ is the chemical potential of the $\alpha$-reservoir. In particular, we set $\mu_\alpha^{\rm eq} = 0$ in thermal equilibrium.

By employing a sum-over-poles expansion for $f^{\sigma}_{\alpha}(\omega)$ and $J_{\alpha \nu s}(\omega)$ in Eq.~\eqref{eqn:FD}, the reservoir correlation functions can be expressed as a linear combination of exponential functions
\begin{equation}
	C_{\alpha \nu s}^{\sigma}(t) = \sum_{p=1}^{P} 
    \eta_{\alpha \nu s p}^{\sigma} e^{-\gamma_{\alpha \nu sp}^{\sigma}t}.
    \label{eqn:Ct-exp-1}
\end{equation}
To be concise, we define the $j$ index as the multi-index $\{\alpha \nu s p\}$ here. We adopt the \Pade spectral decomposition scheme \cite{hu2010communication} to perform the exponential expansion in \Eq{eqn:Ct-exp-1}. 

\begin{figure}[t]
	\includegraphics[width=\columnwidth]{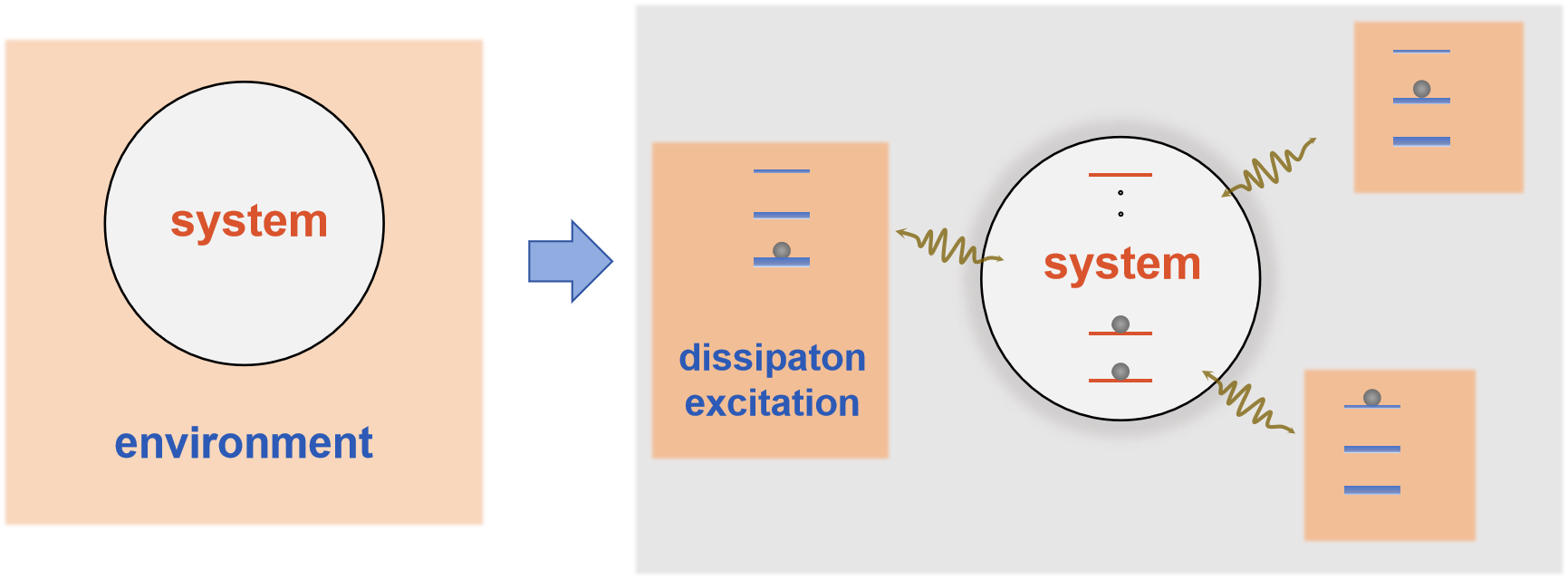}
	\caption{Schematic of the fermionic DQME theory, mapping the original OQS (left) to a dissipaton-embedded system (right). Red and blue bars represent the $N_{_{\rm S}}$ system fermion energy levels and $N_{_{\rm E}}$ memory-carrying dissipaton levels, respectively. 
    Broadening of blue bars indicates each dissipaton's decay rate (inverse lifetime). } \label{fig1}
\end{figure}

In the DQME framework, the expansion of the  reservoir correlation functions defines the dissipatons (quasiparticles of dissipative effects with complex energies) \cite{hu2010communication,Yan14054105,li2024toward}.
$j$ labels the dissipaton levels, and $\sigma$ denotes the dissipaton charge ($\sigma=+$ for hole-type; $\sigma=-$ for electron-type). The imaginary and real parts of $\gamma_j^\sigma$ correspond to the dissipaton energy and inverse lifetime, respectively, while $\eta^\sigma_j$ quantifies the system-dissipaton coupling strength.

Figure~\ref{fig1} illustrates the fermionic DQME theory, where the original OQS is mapped to a dissipaton-embedded system described in terms of the reduced density tensor (RDT), $\bm{\rho} = \rho(\vec{n},\vec{n}';\vec{m}^-,\vec{m}^+)$. Here, vectors $\vec{n}$ and $\vec{n}'$ describe system fermion configurations, while $\vec{m}^-$ and $\vec{m}^+$ represent electron-type and hole-type dissipaton configurations, respectively. The RDT elements with ($\vec{m}^-,\vec{m}^+) = \vec{0}$ yield the system's reduced density operator (RDO). As shown in Fig.~\ref{fig1}, each level can accommodate at most one fermion or dissipaton in accordance with the Pauli exclusion principle. 
In practice, a maximal dissipaton occupation number $M_{\rm max}$ is introduce to facilitate the numerical solution of the DQME \cite{li2024toward}.

The fermionic DQME is given by \cite{li2024toward}:
\begin{align}
        \dot{\bm{\rho}} =\mathcal{L}\bm\rho=& -i[H_{_{\rm S}},\bm{\rho}] - \sum_j \left(
        \gamma_j^- \hat{N}_j \bm{\rho} + \gamma_j^+ \bm{\rho} \hat{N}_j\right) \notag \\
       & -i \sum_j \Big[ \big(\hat{c}_{\nu}^{\dagger} \hat{b}_{j} \bm{\rho} - \hat{b}_{j} \bm{\rho} \hat{c}_{\nu}^{\dagger} \big) + 
       \big(\hat{c}_{\nu} \bm{\rho} \hat{b}^{\dagger}_j -  \bm{\rho} \hat{b}_{j}^{\dagger} \hat{c}_{\nu} \big) \Big] \notag \\
       & -i \sum_j  \left[-\eta_{j}^{-}\, \hat{c}_{\nu} \hat{b}_j^{\dagger} \, \bm{\rho} - (\eta_{j}^{+})^\ast \,\hat{b}_j^{\dagger}\bm{\rho} \hat{c}_{\nu}\right]  \notag \\
       & -i \sum_j \left[ 
       \eta_{j}^{+} \, \hat{c}_{\nu}^{\dagger}\bm{\rho}\, \hat{b}_j
       + (\eta_{j}^{-})^\ast  \bm{\rho}\, \hat{b}_j \hat{c}_{\nu}^{\dagger}
       \right]. \label{eqn:DQME-SQ-1}
\end{align}
Here, $\hat{c}_{\nu}^{\dagger}$ ($\hat{c}_{\nu}$) creates (annihilates) a fermion at the $\nu$-th system fermion level, $\hat{b}_j^{\dagger}$ ($\hat{b}_j$) creates (annihilates) a dissipaton at $j$-th level, and $\hat{N}_j = \hat{b}_j^{\dagger} \hat{b}_j$ gives the dissipaton occupation. 
The RDO of the system, $\rho_0$, is obtained by projecting $\bm\rho$ onto the dissipaton vacuum. 

Fundamentally, dissipatons share similarities with previously proposed pseudomodes \cite{PhysRevA.55.2290,Tam18030402,Lam193721,cirio2023pseudofermion,Lin251289,PhysRevResearch.2.043058}, and the DQME is formally equivalent to the exact HEOM theory \cite{Jin08234703,PhysRevLett.109.266403} for reduced system dynamics. 

\subsection{PINN-DQME method} \label{subsec:pinn}

\begin{figure}[h]
	\includegraphics[width=\columnwidth]{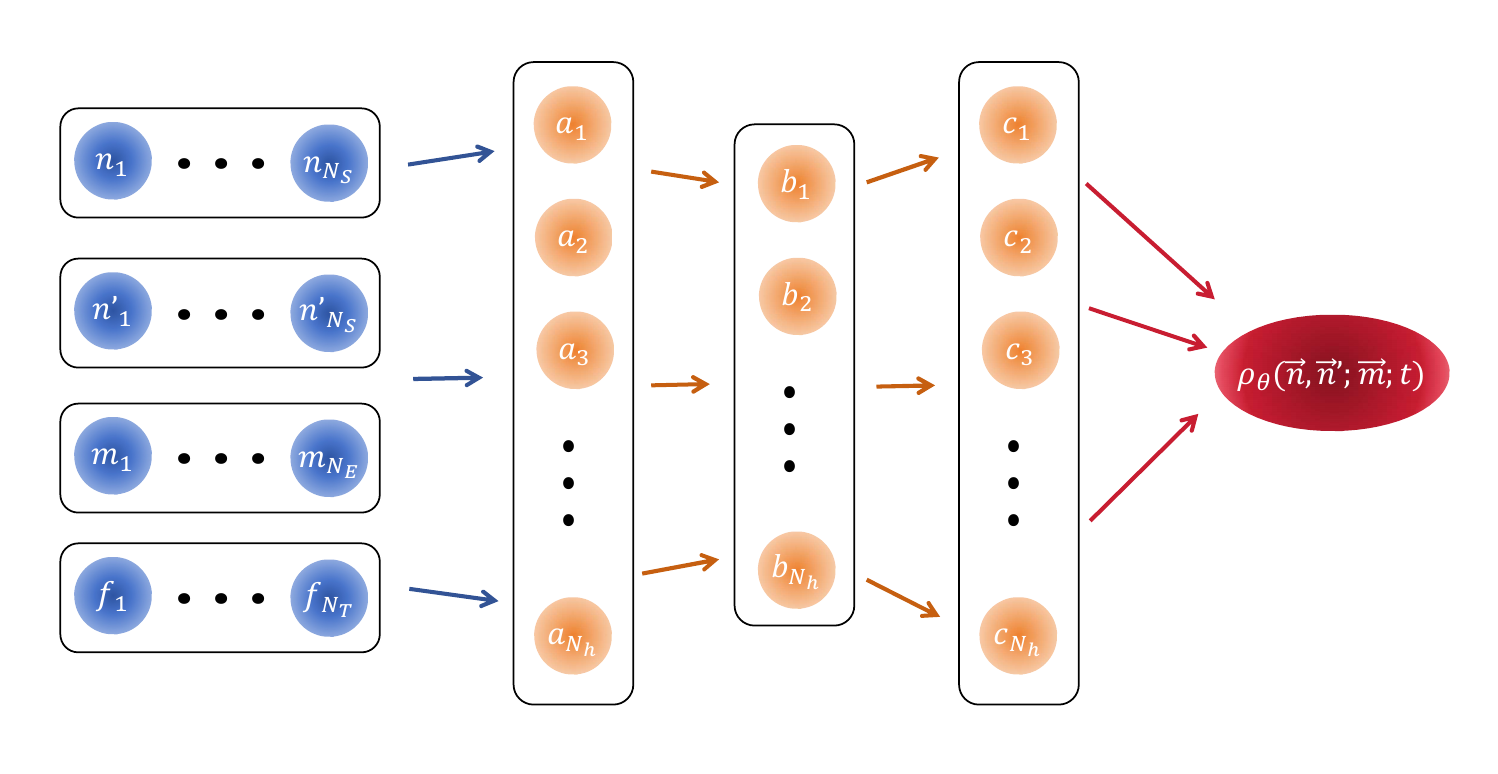}
     \caption{
     Schematic of the neural network representing $\rho_{\rm pre}(\vec{n},\vec{n}';\vec{m};t)$. 
     The input layer consists of three parts: the fermion configurations $\{n_1,\cdots, n_{N_{_{\rm S}}}\}$ and $\{n'_1, \cdots, n'_{N_{_{\rm S}}}\}$ from the $N_{_{\rm S}}$ system levels,
     the dissipaton configurations $\{m_1,\cdots, m_{N_{_{\rm E}}} \}$ from the $N_{_{\rm E}}$ memory-carrying dissipaton levels, and a set of time-varying functions $\{f_1(t),\cdots,f_{N_{_{\rm T}}}(t)\}$. Here, $n_i$ and $m_j$ represent the occupation number on $i$th system level and $j$th dissipaton level, respectively; which take the value of $0$ or $1$. The arrows indicate full connections between layers.  
     }  \label{fig:mlp}
\end{figure}

We utilize a multilayer perceptron (MLP) with complex-valued weights and biases to represent the solution of DQME $\rho(\vec{n},\vec{n}';\vec{m};t)$, where $\vec{m}=(\vec{m}^-,\vec{m}^+)$. The architecture of the MLP is illustrated in Fig.~\ref{fig:mlp}, where the input layer consists of the system nodes ($\vec{n},\vec{n}'$), the environmental nodes $\vec{m}$ and the time nodes $\vec{f}(t)$. The output is the value of the solution projected onto the basis $(\vec{n},\vec{n}';\vec{m})$ at time $t$. Compared to the NQS ansatz, the input layer contains several additional nodes corresponding to the time to encode the information of evolution in the neural network, so this network can represent quantum states at different times rather than at a specific time instant. 
While the conventional PINN methods \cite{raissi2018deep,cuomo2022scientific} usually have only one time node $f_1(t)=t$ in the input layer, here we use a group of time nodes to improve the capability of the MLP for representing complex time evolution.
For an MLP consisting of $K-1$ hidden layers, the computation of its output can be described as a sequence of $K$ matrix-vector multiplications separated by a nonlinear activation function $a:\mathbb{C}\to\mathbb{C}$. Specifically, the RDT of DQME is approximated by
\begin{equation}
    \rho_{\rm pre}(\vec{s};t) \approx F_{K}a(F_{K-1}a(\cdots a (F_{1}(\bst)))).
\end{equation}
Here, $\rho_{\rm pre}(\vec{s};t)$ is a preliminary representation of RDT by the MLP, which is subject to further simplification. $\bst =(\vec{s},\vec{f}(t))$ represents the input of the MLP, $a({\bm x}) = \frac{1}{e^{\bm x}+1}$, and $F_{i}({\bm x}) = {\bm W}_{i}{\bm x}+{\bm b}_{i}$, where ${\bm W}_{i}\in \mathbb{C}^{N_{i}\times N_{i-1}}$ is the complex-valued weight matrix of the $i$-th layer, and ${\bm b}_{i}\in \mathbb{C}^{N_{i}}$ is the complex-valued bias vector of the $i$-th layer. Altogether, ${\bm W}_{i}$ and ${\bm b}_{i}$ form the set of variational parameters $\btheta$ to be determined. 

The dimensionality of the RDT can be reduced by leveraging its inherent symmetry and sparsity \cite{Hou15104112}. First, we enforce the symmetry constraint by constructing
\begin{equation}
    \rho_{\rm sym}(\vec{s};t) = \rho_{\rm pre}(\vec{s};t)
    + (-1)^{\lfloor M^-/2 \rfloor+ \lfloor M^+/2 \rfloor} \rho_{\rm pre}^\ast(\vec{s}^{\, T};t),
    \label{eqn:def-rhosym-1}
\end{equation}
where $\lfloor \cdot \rfloor$ denotes the floor function, $M^{\sigma} = \sum_j m_j^{\sigma}$, $m_j^{\sigma}$ is the $j$-th element of $\vec{m}^\sigma$, and $\rho_{\rm pre}(\vec{s};t)$ is the MLP representation of the solutions. Here, $\vec{s} \equiv (\vec{n},\vec{n}';\vec{m}^-,\vec{m}^+)$ represents a visible state, while $\vec{s}^{\, T} \equiv (\vec{n}',\vec{n};\vec{m}^+,\vec{m}^-)$ is its block-swapped counterpart.
Next, we introduce a filter function $\mathcal{F}_{\rm{spa}}(\vec{s})$ that identifies, {\it a priori}, the zero elements of $\bm\rho$ based on the sparsity pattern dictated by the forms of $\Hs$ and $\Hse$ \cite{Hou15104112}, which keeps invariant during the evolution.
In practice, $\mathcal{F}_{\rm{spa}}(\vec{s})$ normally reduces the $\vec{s}$-space by over an order of magnitude. The resulting symmetrized and filtered RDT is given by $\rho(\vec{s};t) = \mathcal{F}_{\rm spa}(\vec{s}) \rho_{\rm sym} (\vec{s};t)$ \cite{cao2026simulating}.

Within the framework of PINN, the time evolution of the RDT is obtained by constructing and minimizing a loss function $L$, which consists of three components:
\begin{align}
	L &= \omega_{\rm R}L_{\rm R} + \omega_{\rm I} L_{\rm I} +\omega_{\rm tr}L_{\rm tr}, \\\
	L_{\rm R} &=
	\sum_{i=0}^{N_{\rm R}}\frac{\Vert\dot{\bm{\rho}_{\theta}}(\tau_i)-\mathcal{L}\bm\rho_{\theta}(\tau_i)\Vert_2^2} {[{\rm tr}_{_{\rm S}}(\rho_0(\tau_i))]^2}, \\ 
	L_{\rm I} &= \Vert\bm\rho_{\theta}(\tau_0)-\bm\rho(\tau_0)\Vert_M^2,\label{eqn:initial1}\\
    L_{\rm tr} &=
	\sum_{i=1}^{N_{\rm R}} \vert {\rm tr}_{_{\rm S}}(\rho_{0}(\tau_i))-1\vert^2,
\end{align}
where $L_{\rm R}$ is the residual term from DQME, $L_{\rm I}$ comes from the initial condition $\bm\rho_{\theta}(\tau_0)=\bm\rho(\tau_0)$,  $L_{\rm {\rm tr}}$ is introduced to enforce the physical constraints that the trace of RDO equals 1, and $\omega_{\rm R}$, $\omega_{\rm I}$ and $\omega_{\rm tr}$ are the weights assigned to the equation residual, the initial condition, and the trace residual, respectively. In practice, we empirically choose  $\omega_{\rm R}=0.2$, $\omega_{\rm I}=0.8$ and $\omega_{\rm tr}=20$, where not otherwise specified. In these terms, $N_{\rm R}$ is the count of residual points (indexed by $\tau_i$), $\bm\rho_{\theta}(t)$ is the parametric RDT at time $t$, $\rho_0(t)$ is the parametric RDO of the system at time $t$, and ${\rm tr}_{_{\rm S}}$ means taking the trace of the system subspace. The time derivative of the RDT, $\dot{\bm \rho}$, is computed by numerical differentiation with a sufficiently small time difference of $\delta t=1.5\times10^{-9}\,\Gamma^{-1}$. $\Vert \cdot\Vert_2$ denotes the 2-norm, while $\Vert\cdot \Vert_M^2$ is defined as: $\Vert\bm \rho\Vert_M^2 = \sum_{\vec{s}}e^{\lambda M(\vec{s})}\vert\rho(\vec{s})\vert^2$ with $M(\vec{s})=\sum_jm_j$ and the empirical exponent $\lambda=-3$. This exponent is introduced to emphasize the importance of elements with low $M$ in RDT, which hold greater relevance to physical observables. 
Minimizing this loss function, we can find the possible approximation of the solutions of DQME under specific initial conditions.

In cases where the quantum state of the system undergoes significant changes over the studied time scale, the complex dynamics can render the full evolution difficult to simulate with a single neural network \cite{krishnapriyan2021characterizing}, due to the rugged loss landscape arising from the governing equations. To address this, we adopt a domain decomposition strategy \cite{jagtap2020conservative,krishnapriyan2021characterizing}, which divides the entire time interval into several subdomains and approximates the evolution within each using an individual neural network. The loss function for the $p$-th subdomain is defined as:
\begin{align}
	L(\theta_p) &= \omega_{{\rm R}}L_{{\rm R}_p} + \omega_{{\rm I}} L_{{\rm I}_p} + \omega_{{\rm tr}} L_{{\rm tr}_p},\\
	L_{{\rm R}_p} &=
	\sum_{i=0}^{N_{{\rm R}_p}}\frac{\Vert\dot{\bm{\rho}}_{\theta_{p}}(\tau_{i,p})-\mathcal{L}\bm\rho_{\theta_{p}}(\tau_{i,p})\Vert_2^2}{{[{\rm tr}_{_{\rm S}}(\rho_0(\tau_{i,p}))]^2}},\\  
	L_{{\rm I}_p} &= \Vert\bm\rho_{\theta_p}(t_{p-1})-{\bm\rho}_{\theta_{p-1}}(t_{p-1})\Vert_M^2,\label{eqn:initial2} \\
    L_{{\rm tr}_p}&=\sum_{i=0}^{N_{{\rm R}_p}} \vert {\rm tr}_{_{\rm S}}(\rho_{0}(\tau_{i,p}))-1\vert^2.
\end{align}
Here, $\omega_{{\rm R}}=0.2$, $\omega_{{\rm I}}=0.8$ and $\omega_{{\rm tr}}=20$ are empirical weights for the residual, initial-condition, and trace-preserving terms, respectively. The terms $L_{{\rm R}_p}$, $L_{{\rm I}_p}$ and $L_{{\rm tr}_p}$ denote the residual loss, initial-condition loss, and trace-preserving loss. The set $\{\tau_{i,p}\}$ represents the residual points, with $N_{{\rm R}_p}$ being their count in the $p$-th subdomain, and $\bm\rho_{\theta_p}$ is the parameterized RDT for this subdomain. The initial time for the $p$-th subdomain is $t_{p-1} = \tau_{0,p}$ (with $t_0=0$). Specifically, the initial-condition loss $L_{{\rm I}_p}$ in Eq.~\eqref{eqn:initial2} enforces continuity by using the neural network output optimized in the $(p-1)$-th subdomain as the initial condition for the $p$-th subdomain.

To efficiently optimize the PINN, we employ a warm‑start strategy, where the neural‑network parameters obtained from the preceding subdomain serve as the initial guess for the current subdomain. Compared to a random initialization, this approach typically reduces the initial‑condition loss by several orders of magnitude, thereby significantly accelerating the overall optimization.

There are many methods in machine learning to optimize the loss function, such as stochastic gradient descent (SGD) \cite{robbins1951stochastic}, Adam \cite{kingma2014adam}, and AdamW \cite{loshchilov2017decoupled}. The most common optimization method in PINN is limited-memory
Broyden-Fletcher-Goldfarb-Shanno algorithm (L-BFGS) \cite{cuomo2022scientific}. However, all of them failed when applied to the PINN-DQME method due to the significant complexity of DQME. Consequently, we chose the standard BFGS algorithm  \cite{fletcher2013practical} , which is 
highly accurate but somewhat expensive, to minimize the loss function involved in the PINN-DQME method.

\section{Results and discussion}  \label{sec:result}

\subsection{Simulation of non-Markovian dissipative dynamics of Anderson impurity model}

\begin{figure}[t]
	\includegraphics[width=\columnwidth]{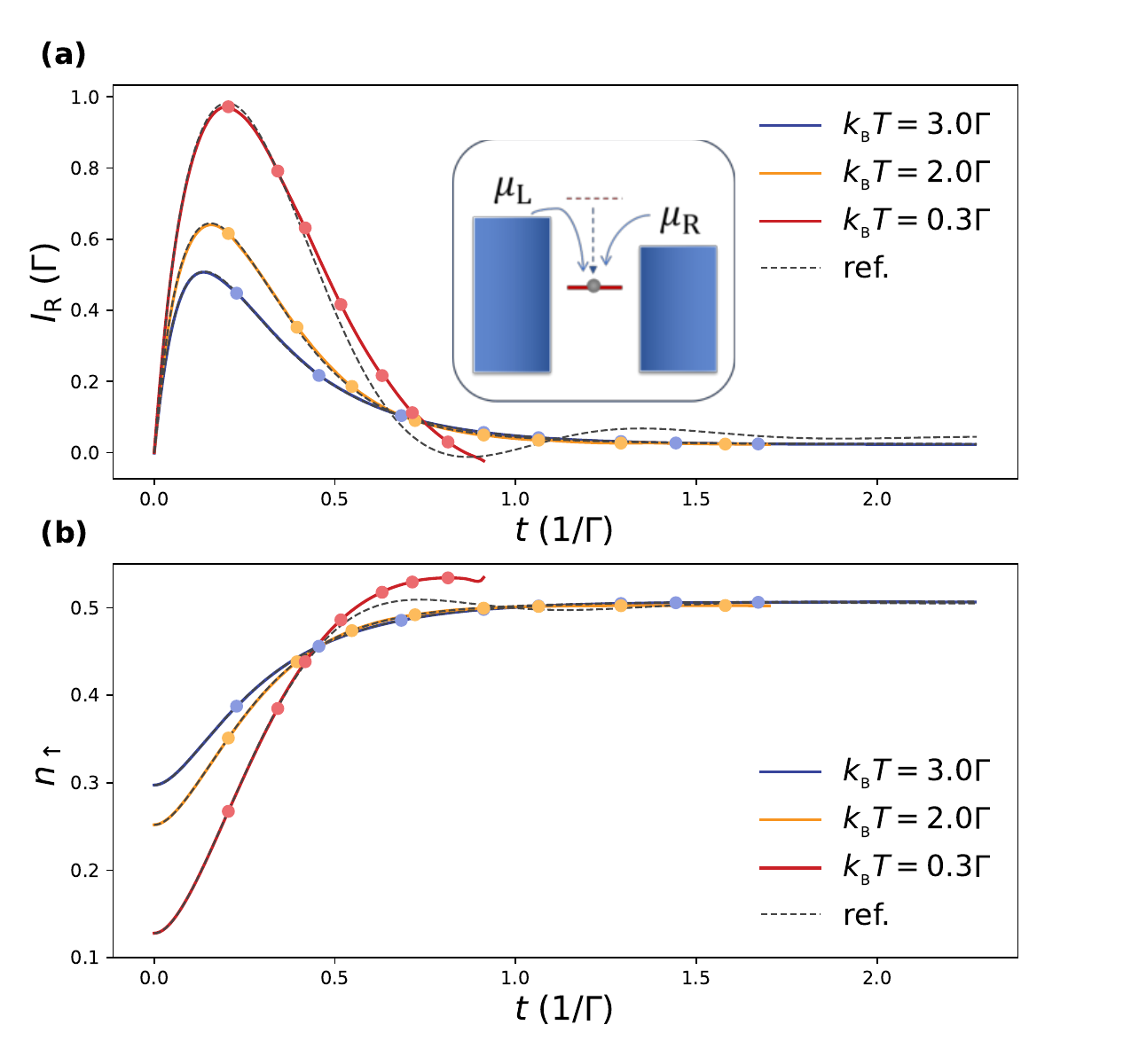}
     \caption{
      (a) The evolution of the electric current flowing into the impurity from the right reservoir at different temperatures. Inset: schematic of quantum dissipative dynamics for an impurity coupled to left (L) and right (R) reservoirs with chemical potentials $\mu_{\rm L}$ and $\mu_{\rm R}$. (b) The evolution of the occupation number of the up spin electron in the impurity at two different temperatures. 
      The purple and red dots represent the decomposition points $\{t_p\}$ for the time domain. 
      The results of PINN-DQME (solid lines) are benchmarked against the reference values (dashed lines) obtained by the HEOM method.       
      For the case of $k_{\rm B}T=0.3\,\Gamma$, the PINN-DQME calculation is terminated at $t_{\rm end} = 0.82\,\Gamma^{-1}$, as the minimization of loss function becomes increasingly demanding; see the main text for details.
      The MLP hyperparameters are chosen as $K=4$ and $N_{\rm h}=35$.
      For the case of $k_{_{\rm B}}T=3.0\,\Gamma$, the mapping functions $f_1(t)=t$ and $f_2(t)=f_3(t)=0$ are used in all subdomains.
      For the case of $k_{_{\rm B}}T=2.0\,\Gamma$, the mapping functions       
      $f_1(t)=t$, $f_2(t)=t^2$, and $f_3(t)=t^3$ are used in the first three subdomains, and $f_1(t)=t$, $f_2(t)=t^{1.5}$, and $f_3(t)=t^{0.5}/(t+0.015\Gamma^{-1})$ are used in the later subdomains.
      For the case of $k_{_{\rm B}}T=0.3\,\Gamma$, the mapping functions       
      $f_1(t)=t$, $f_2(t)=t^2$, and $f_3(t)=t^3$ are used in the first two subdomains, and $f_1(t)=t$, $f_2(t)=t^{1.5}$, and $f_3(t)=t^{0.5}/(t+0.015\Gamma^{-1})$ are used in the later subdomains.
      System energetic parameters are chosen as (in units of $\Gamma$): $\epsilon_0 = U_0/2 = 2$, $\Delta\epsilon = -7$, and $\Delta U = 6$.}  \label{fig:results}
\end{figure}

We consider an OQS comprising a localized impurity symmetrically coupled to two noninteracting electron reservoirs. This setup is important for understanding electron transport through molecular junctions. The OQS is described by the single-impurity Anderson model \cite{And6141}. The impurity Hamiltonian is
$
H_{_{\rm S}}(t) = \epsilon_{0} (\hat{n}_\uparrow + \hat{n}_\downarrow)  + U_0\hat{n}_{\uparrow}\hat{n}_{\downarrow} + 
{\Theta}(t-t_0) [\Delta \epsilon (\hat{n}_\uparrow + \hat{n}_\downarrow) + \Delta U\hat{n}_{\uparrow}\hat{n}_{\downarrow}]
$,
where $\hat{n}_s$ is the occupation operator for spin-$s$ electrons, $\epsilon_0$ is the impurity energy, and $U_0$ is the Coulomb interaction energy.
At time $t_0 = 0$, a sudden change shifts the impurity's parameters by $\Delta \epsilon$ and $\Delta U$, and a bias voltage is simultaneously switched on, establishing a chemical potential difference between the reservoirs. 
The inset of Fig.~\ref{fig:results}(a) depicts the resulting  open quantum dynamics. The impurity level shift triggers electron transfer from the reservoirs to the impurity on a relatively short timescale. Eventually a steady electric current flows through the impurity. At sufficiently low reservoir temperatures, environmental memory effects influence the electron transition process, resulting in non-Markovian oscillations.
In the following, we also calculate the non-Markovian open quantum dynamics using the fermionic HEOM method implemented in the HEOM-QUICK program \cite{ye2016heom,zhang2024heom}, and the numerical outcomes are served as reference values for assessing the accuracy of the PINN-DQME method.


Figure~\ref{fig:results}(a) and (b) exhibit the time-dependent current through the right reservoir $I_{\rm R}(t)$ and the occupation number of spin-up electrons on the impurity level $n_{\uparrow}(t)$, respectively, at high ($k_{\rm B}T=3.0\,\Gamma$), intermediate ($k_{\rm B}T=2.0\,\Gamma$) and low ($k_{\rm B}T=0.3\,\Gamma$) temperatures. 
The time-domain decomposition is illustrated through the dots on the lines. Within each subdomain, the time evolution of RDT is quantified by an MLP with three hidden layers ($K=4$), with each hidden layer comprised of 35 nodes ($N_{\rm h}=35$). The residual points (incremental time steps) are taken at intervals of $\Delta \tau \approx 0.015\,\Gamma^{-1}$. In the PINN-DQME framework, the expectation value of an observable $\hat{X}$, $X = \la \hat{X} \ra$, is evaluated from the parametrized RDT $\bm\rho_{\theta}$.

For the cases of $k_{\rm B}T=3.0\,\Gamma$ and $k_{\rm B}T=2.0\,\Gamma$, the PINN-DQME method yields highly accurate results that agree excellently with the reference data throughout the entire time domain, as evidenced by the rather small relative time-integrated errors, i.e., $\mathcal{E}_X < 1.7\%$ with $X = I_{\rm R}$ and $n_{\uparrow}$, where the relative errors are defined by
\begin{equation}  \label{eqn:def-err}
 \mathcal{E}_X \equiv \frac{\int_{t_0}^{t_{{\rm end}}} \lvert X(\tau) - X^{{{\rm ref}}}(\tau) \rvert \, {\rm d}\tau} {\int_{t_0}^{t_{{\rm end}}} \frac{1}{2} \big( \lvert X(\tau) \rvert + X^{{{\rm ref}}}(\tau) \rvert  \big) {\rm d} \tau}.
\end{equation} 
Here, $X^{{{\rm ref}}}$ denotes the reference values calculated by the HEOM method, and $t_{\rm end}$ is a chosen end point for time evolution. Particularly, the MLP trained within a subdomain $[1.44\,\Gamma^{-1},1.67\,\Gamma^{-1}]$ at $k_{\rm B}T=3.0\,\Gamma$ can be straightforwardly extended to quantify the quantum dissipative dynamics up to the end time of $t_{\rm end} = 2.3\,\Gamma^{-1}$ without changing any of its parameters. This highlights the remarkable practicality of PINN for capturing open quantum dynamics where non-Markovian effects are less prominent, because the dissipatons' lifetimes are shorter at higher temperatures \cite{li2023dissipatons,li2024quantum}.  



In contrast, for the case of $k_{\rm B}T=0.3\,\Gamma$, the performance of PINN deteriorates significantly as the open quantum dynamics becomes strongly non-Markovian. As shown in Fig.~\ref{fig:results}, while the results of PINN-DQME closely match the reference values in the first two subdomains, they begin to deviate noticably from the third subdomain onward. Moreover, the minimization of loss function becomes increasingly difficult in later subdomains.  Consequently, the PINN-DQME calculation undergoes an early termination at $t_{\rm end} = 0.82\,\Gamma^{-1}$.

\begin{figure}[t]
\includegraphics[width=\columnwidth]{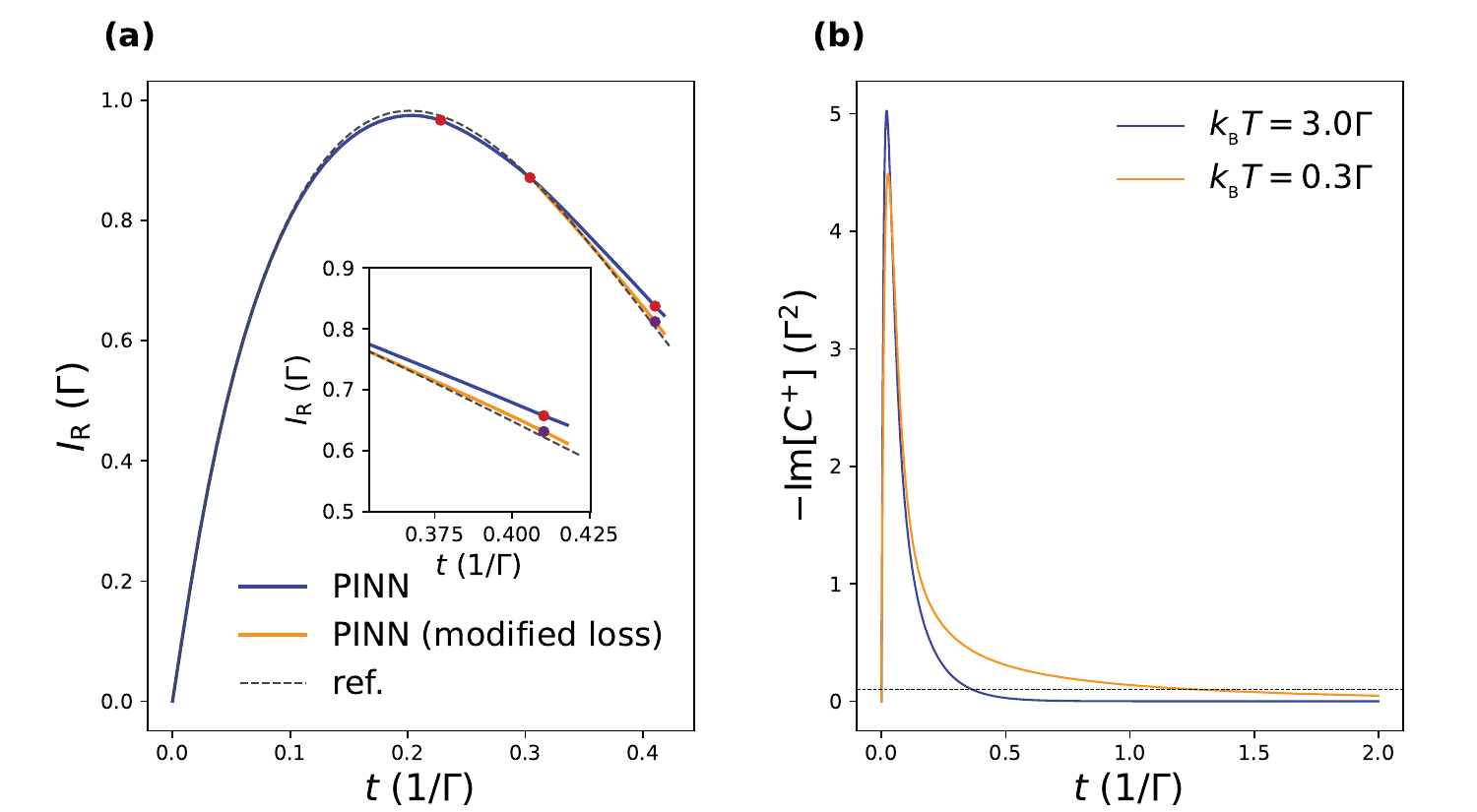}
     \caption{
     (a) Comparison of electric current for two loss functions for the case of $k_{\rm B}T=0.3\,\Gamma$. The main panel shows the time evolution of the electric current obtained with the original loss function (blue line) and the modified loss function (orange line); see the main text for details. Purple and red dots mark the domain‑decomposition points
     $t_p$. The inset provides a zoomed‑in view of the third subdomain $[0.3\,\Gamma^{-1}, 0.4\,\Gamma^{-1}]$, where the difference between the two loss functions becomes more apparent.
     (b) Imaginary part of the hybrid correlation function $C^{+}(t)$ at different temperatures. The dashed horizontal line is at $y=10^{-1}\,\Gamma^2$.
}  \label{fig:error_cumulation}
\end{figure}

The considerable deviation observed in the low temperature case is a manifestation of error accumulation during the time evolution, which originates from the MLP's inability to fully represent the quantitative non-Markovian memory effects within a given time subdomain. To elaborate, we modify the loss function in \Eq{eqn:initial2} 
that encodes the initial condition to $L_{{\rm I}_p} = \Vert\bm\rho_{\theta_p}(t_{p-1})-{\bm\rho}^{{\rm ref}}(t_{p-1})\Vert_M^2$ for the third subdomain $[0.3\,\Gamma^{-1}, 0.4\,\Gamma^{-1}]$. Here, the reference values derived from the HEOM calculation ${\bm\rho}^{{\rm ref}}(t_{p-1})$ replaces the MLP-approximated counterpart $\bm\rho_{\theta_{p-1}}(t_{p-1})$. As demonstrated in Fig.~\ref{fig:error_cumulation}(a), this adjustment yields substantially improved accuracy in the third subdomain. The result  underscores the importance of maintaining high accuracy at subdomain boundaries. However, such a requirement can be highly demanding given the limited representational power of the MLP adopted in this study.

This error accumulation primarily stems from the strongly non‑Markovian memory at low temperatures. Non‑Markovianity describes the backflow of information from the environment to the system. Within the PINN framework, this backflow leads to complex time-dependent evolution that is more prone to error accumulation. Consequently, when the environment exhibits stronger non‑Markovian memory, initial errors tend to persist and propagate over longer time intervals, exerting a prolonged influence on the subsequent temporal evolution. Such error accumulation is particularly detrimental to PINN‑based simulations. Figure~\ref{fig:error_cumulation}(b) illustrates the imaginary part of the hybrid correlation function $C^{+}(t)$ at different temperatures. The dashed horizontal line marks $y=10^{-1}\,\Gamma^2$. The correlation function at $k_{\rm B}T=0.3\,\Gamma$ decays significantly more slowly than that at $k_{\rm B}T=3.0\,\Gamma$, indicating stronger non‑Markovian effects at lower temperatures.

\begin{table}[t]
\centering
\caption{Comparison of parameter counts for the PINN-DQME and NQS-DQME methods.}  \label{tab:nqs}
\begin{tabular*}{\columnwidth}{@{\extracolsep\fill}lcccccc}
\hline\hline
$k_{\rm B}T/\Gamma$  & $3.0$ & $2.0$ & $0.3$  \\
\hline
$N_{\rm para\,(PINN)}$ per subdomain  & 6182  &8472  & 10152   \\
$N_{\rm para\,(NQS)}$ per time point & 3712 &4584  & 10510  \\
\hline\hline
\end{tabular*}
\end{table}

Then, we compare the computational cost of the PINN-DQME method and the NQS-DQME method \cite{cao2026simulating}.
We first compare parameter counts in Table~\ref{tab:nqs}. PINN‑DQME uses similar parameters per subdomain as NQS‑DQME per time point, but requires only $\sim$10 subdomains for the full evolution versus $\sim$1000 time points for NQS‑DQME, resulting in a much smaller total. A direct fair runtime comparison is challenging because the two approaches are implemented in different frameworks, and their convergence criteria, architectures, and hyperparameters differ significantly, which could make a simple runtime comparison misleading. Nevertheless, a partial quantitative comparison is still possible: the time for PINN to reduce the loss from a suitable initialization to $10^{-4}$ is of the same order as the time for NQS to evolve through the same subdomain. This suggests that PINN‑DQME has the potential to outperform NQS‑DQME in weakly non‑Markovian regimes where only moderate precision is required.

\subsection{Enhancing the performance of PINN-DQME}

In the following we explore several aspects that have significant impact on the performance of PINN-DQME. These include the choice of the residual points, the feature mapping functions for the time nodes, etc.

The selection of residual points substantially influences both the accuracy of the results and the computational cost of training. Excessively dense residual points lead to high resource consumption, whereas overly sparse points cause significant deviations from the reference data. To balance efficiency and precision, we adopt a practical training strategy based on adaptive residual-point sampling: beginning with a relatively sparse initial distribution and progressively increasing the point density in stages.

\begin{figure}[t]
 \includegraphics[width=\columnwidth]{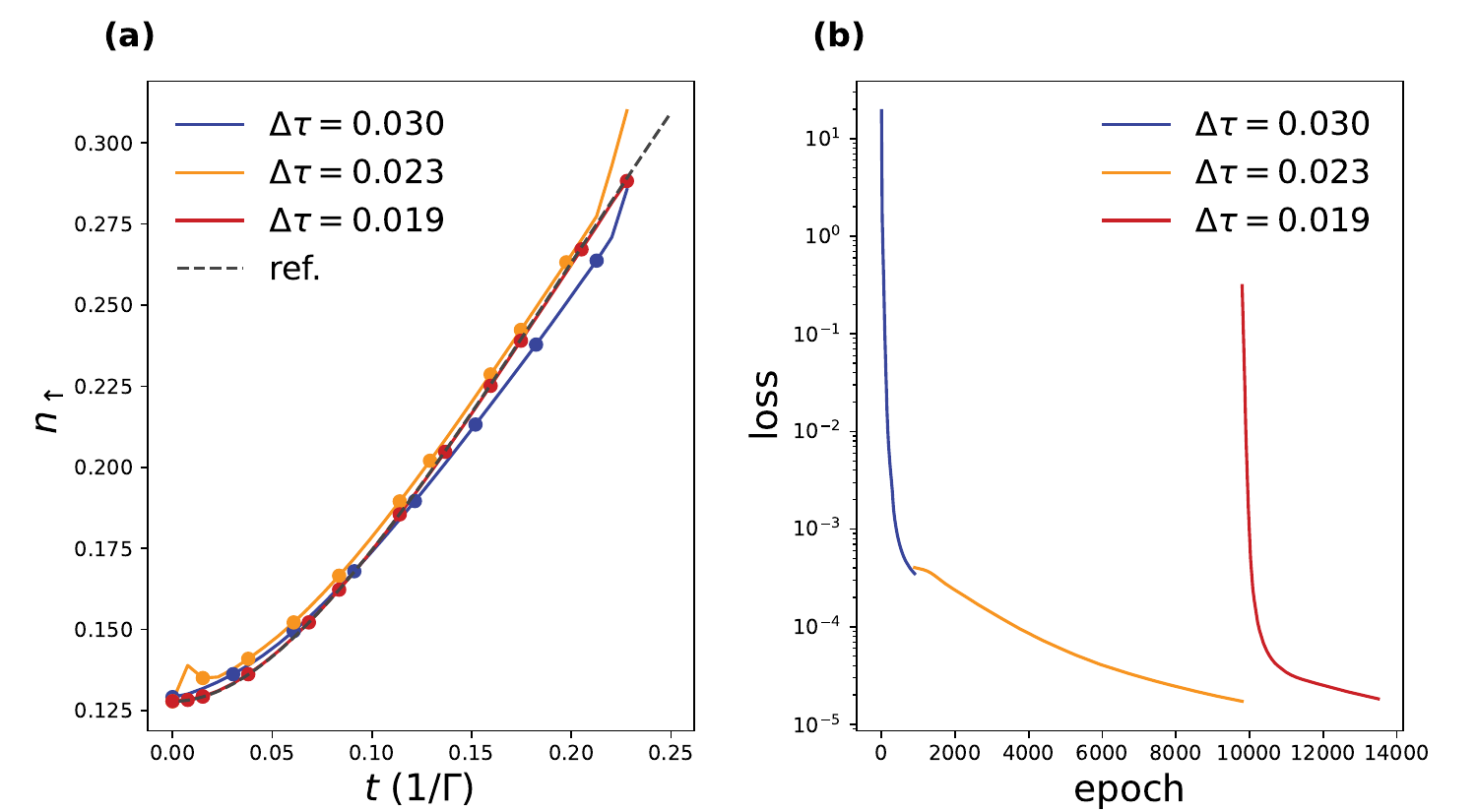}
     \caption{Staged training process with adaptive residual-point sampling in the first subdomain $[0, 0.228\, \Gamma^{-1}]$ for the case of $k_{\rm B}T=0.3\,\Gamma$. 
     (a) Evolution of the population $n_\uparrow$ during training. Solid lines show the PINN-DQME output at different stages, corresponding to successively reduced average residual-point spacings $\Delta\tau$. The dots on each curve mark the residual points used in that stage. The dashed black line represents reference data. (b) Corresponding decay of the training loss (log scale) for each stage. The colored curves track the loss progression for the $\Delta\tau$ values shown in (a), demonstrating stable convergence throughout the adaptive refinement.
     }  \label{fig:training}
\end{figure}

Figure~\ref{fig:training} illustrates this staged training procedure within the first subdomain $[0, 0.228\,\Gamma^{-1}]$ for the case of $k_{\rm B}T=0.3\,\Gamma$. Here, $\Delta\tau$ denotes the average spacing between adjacent residual points, which are held fixed within each sub-optimization stage and then refined in the next.
In Fig.~\ref{fig:training}(a), the blue curve corresponds to the coarsest initial distribution ($\Delta\tau = 0.03\,\Gamma^{-1}$). After optimization until the loss falls below $1 \times 10^{-3}$,  the residual points are refined to a smaller spacing ($\Delta\tau = 0.023\,\Gamma^{-1}$, yellow curve). Because the network output varies smoothly around the newly added points, the loss does not exhibit a sharp jump and continues to decrease to about $2 \times 10^{-5}$.
However, when examining the neural network output on a finer temporal grid, a non‑smooth region (visible as a cusp) emerges in the yellow curve near the starting point of the subdomain. To address this, the point density is further increased ($\Delta\tau = 0.019\,\Gamma^{-1}$, red curve), with additional residual points placed specifically around the cusp region. The introduction of these points initially raises the loss, as seen at the start of the red curve in Fig.~\ref{fig:training}(b), but continued optimization drives the loss back down to about $2 \times 10^{-5}$. Consequently, the PINN‑DQME result (red curve) aligns closely with the reference data (dashed line). The overall decay of the training loss across all stages is displayed on a logarithmic scale in Fig.~\ref{fig:training}(b), where the colors distinguish the different residual‑point configurations. The plot confirms that loss reduction remains stable throughout the adaptive refinement process, validating the effectiveness of the proposed staged training strategy.



\begin{figure}[h]
	\includegraphics[width=\columnwidth]{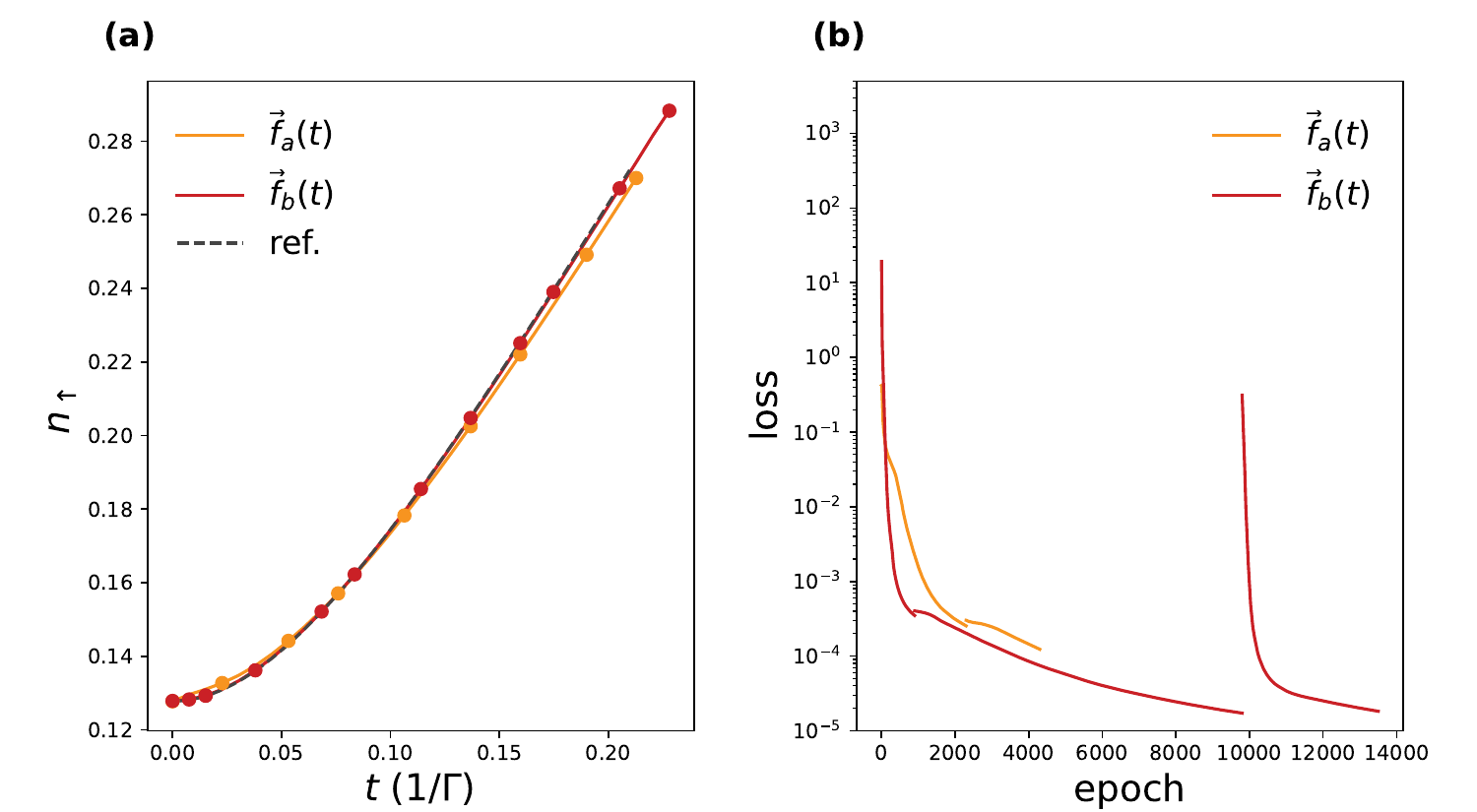}
     \caption{Comparison of PINN-DQME performance using different input feature mappings for the time nodes in the first subdomain $[0.0,0.228\,\Gamma^{-1}]$ at $k_{\rm B}T = 0.3\,\Gamma$.
     (a) Population dynamics $n_\uparrow(t)$. Solid lines show the results obtained using the linear mapping $\vec{f}_a(t): \{t\}$ (orange) and the polynomial mapping $\vec{f}_b(t):\{t, t^2, t^3\}$ (red). The black dashed line is the reference data. Dots on each curve mark the residual points used during optimization of PINN. (b) Corresponding training loss (log scale) versus epoch. The loss for $\vec{f}_b(t)$ (red) decays faster and to a lower final value than that for $\vec{f}_a(t)$ (orange), consistent with the superior accuracy shown in (a).
     }  \label{fig:ft}
\end{figure}

Next, we investigate how the choice of the feature mapping functions $\vec{f}(t)$ for the time nodes in the input layer affects the performance of PINN-DQME. Figure~\ref{fig:ft} compares results obtained using two different mappings within the first subdomain $[0, 0.228\,\Gamma^{-1}]$ for the case of $k_{\rm B}T=0.3\,\Gamma$: a simple linear mapping 
$\vec{f}_a(t): \{t\}$, and a richer polynomial mapping $\vec{f}_b(t):\{t, t^2, t^3\}$. As depicted in Fig.~\ref{fig:ft}(a), the richer feature set $\vec{f}_b(t)$ yields significantly more accurate population dynamics than the linear mapping $\vec{f}_a(t)$. This improvement is corroborated by the loss curves in Fig.~\ref{fig:ft}(b), where $\vec{f}_b(t)$ achieves a lower final loss, confirming that enhanced input features facilitate both faster convergence and higher accuracy. Here, the time-feature mapping functions were chosen based on simplicity and the physical behavior of the system. Specifically, $\vec{f}_a(t) = \{t\}$ and the richer polynomial mapping $\vec{f}_b(t) = \{t, t^2, t^3\}$ were selected for their simplicity, although other mappings are also possible. However, in our tests, exponential and logarithmic mappings (e.g. $\{e^{t},\ \log t\}$) did not perform as well as the polynomial ones. For later subdomains, the term $t^{0.5}/(t + 0.015\Gamma^{-1})$  was designed to capture the decreasing trend of the current. Its numerator $t^{0.5}$ ensures that the term vanishes at $t=0$, similar to the polynomial mappings, while the constant $0.015\,\Gamma^{-1}$ in the denominator regularizes the expression, preventing divergence and ensuring a finite value at $t=0$.



We have also explored several other techniques to enhance PINN‑DQME performance, such as replacing the simple MLP with a ResNet architecture \cite{he2016deep,cheng2021deep} and modifying the initial‑condition loss to incorporate multiple residual points at subdomain boundaries. While these modifications yield modest accuracy improvements, they remain insufficient to reliably capture the full time evolution at $k_{\rm B} T = 0.3\,\Gamma$. These results highlight the inherent limitations of the current PINN framework in representing complex non‑Markovian open quantum dynamics.

\section{Concluding remarks} \label{sec:conclude}



In this work, we have integrated the PINN approach into the NQS framework to simulate open quantum system dynamics. This PINN-DQME method circumvents the computationally expensive TDVP and demonstrates high accuracy in simulating high-temperature Markovian dynamics. However, it encounters challenges with accumulating numerical error when applied to complex non-Markovian regimes. This limitation suggests that the domain decomposition strategy, while effective for many classical PDEs, requires further refinement and adaptation for intricate quantum dynamical settings.

The precise and efficient characterization of non-Markovian memory effects is often regarded as an ``ultimate toughness test'' \cite{Schuld19} for numerical simulations of open quantum dynamics. Compared to recent advances in pure NQS approaches \cite{cao2026simulating}, the current implementation of PINN-DQME adopts a more radical architecture: instead of representing the wave function at a single time point, a single neural network is optimized to represent the temporal evolution over a finite time interval.
This approach offers potential gains in both speed and fidelity.
However, this comes at the cost of a significantly more complex global optimization problem, necessitating the use of the accurate but computationally intensive BFGS algorithm and leading to reduced precision when simulating intricate non-Markovian dynamics. Nevertheless, its success in the high-temperature regime suggests that this direction is worthy of further exploration. The accuracy shortfall primarily stems from error accumulation in the time evolution, while the speed disadvantage is related to the use of a relatively slow BFGS optimizer. Future work may address these aspects by employing more powerful network architectures, such as transformers, or by incorporating more efficient training schemes.

Although the present framework does not fully overcome the profound difficulty of simulating strong non-Markovian dynamics, it establishes a new paradigm for synergistically combining artificial neural networks with quantum dynamical equations. It thus opens a promising alternative pathway for the simulation of open quantum system dynamics, inviting further exploration at the intersection of machine learning and chemical physics.

\begin{acknowledgements}

Support from the National Natural Science Foundation of China (Grants No.\ 22393912, No.\ 22425301, No.\ 22373091 and No.\ 22573099), the Quantum Science and Technology--National Science and Technology Major Project (Grants No.\ 2021ZD0303301 and No.\ 2021ZD0303306), the AI for Science Foundation of Fudan University (Grant No. FudanX24AI023), and the Strategic Priority Research Program of Chinese Academy of Sciences (Grant No.\ XDB0450101) is gratefully acknowledged. 
\end{acknowledgements}

\section*{Conflict of interest}

The authors have no conflicts to disclose.

\section*{Data availability}
The raw and processed data required to reproduce these findings 
and the source code for the numerical solver developed in this work are available at \url{https://github.com/caolong-cn/NQS-DQME_fermions}.

\begin{figure}[t]
	\includegraphics[width=\columnwidth]{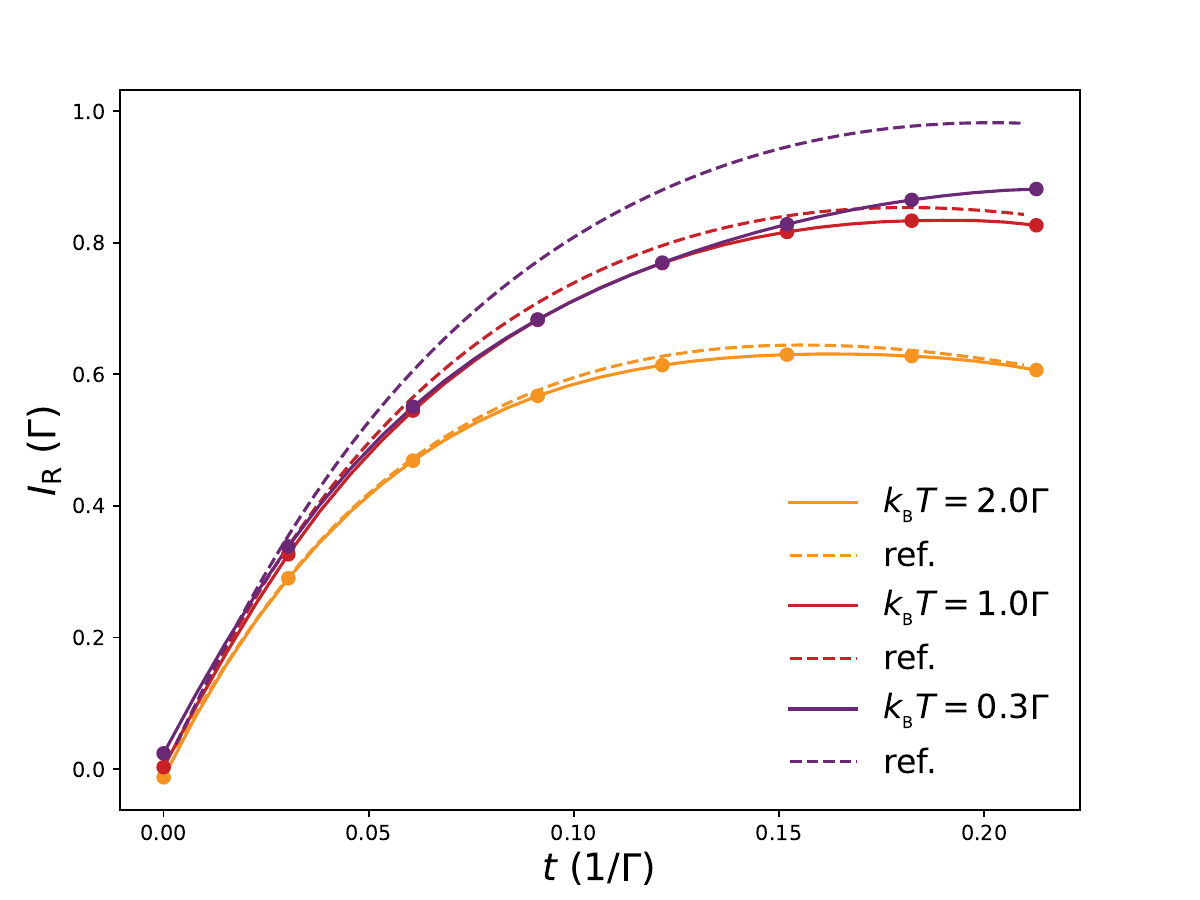}
	\caption{
    Comparison of the electric current at different temperatures 
    in the first subdomain $[0, 0.228\, \Gamma^{-1}]$ with a similar loss function value ($\sim 3\times10^{-4}$). 
    The mapping functions are $f_1(t)=t$, $f_2(t)=t^2$, and $f_3(t)=t^3$ for all cases. Dots on each curve mark the residual points used during optimization of PINN.
 } \label{fig:comparisonT}
\end{figure}

\appendix
\section{Exploring the reliable temperature regime of PINN-DQME method} \label{sec:comparisonT}

Figure~\ref{fig:comparisonT} compares the electric current at different temperatures in the first subdomain $[0, 0.228\, \Gamma^{-1}]$ while maintaining a similar loss function value ($\sim 3\times10^{-4}$). It is observed that, as the temperature decreases, the deviation from reference values increases significantly, despite the comparable loss. This growing deviation reflects the heightened precision requirement at lower temperatures due to the enhanced non-Markovianity.

In our numerical tests, we find that reducing the loss function below $10^{-4}$ is difficult to achieve. As demonstrated in Fig.~\ref{fig:comparisonT}, the precision required for temperatures below $\Gamma$ is less than $10^{-4}$. Therefore, the reliable temperature regime of the present PINN-DQME method is $k_{\rm B} T>\Gamma$.

\section{Comparison of different optimizers}  \label{sec:comparisonOp}

\begin{figure}[t]
	\includegraphics[width=\columnwidth]{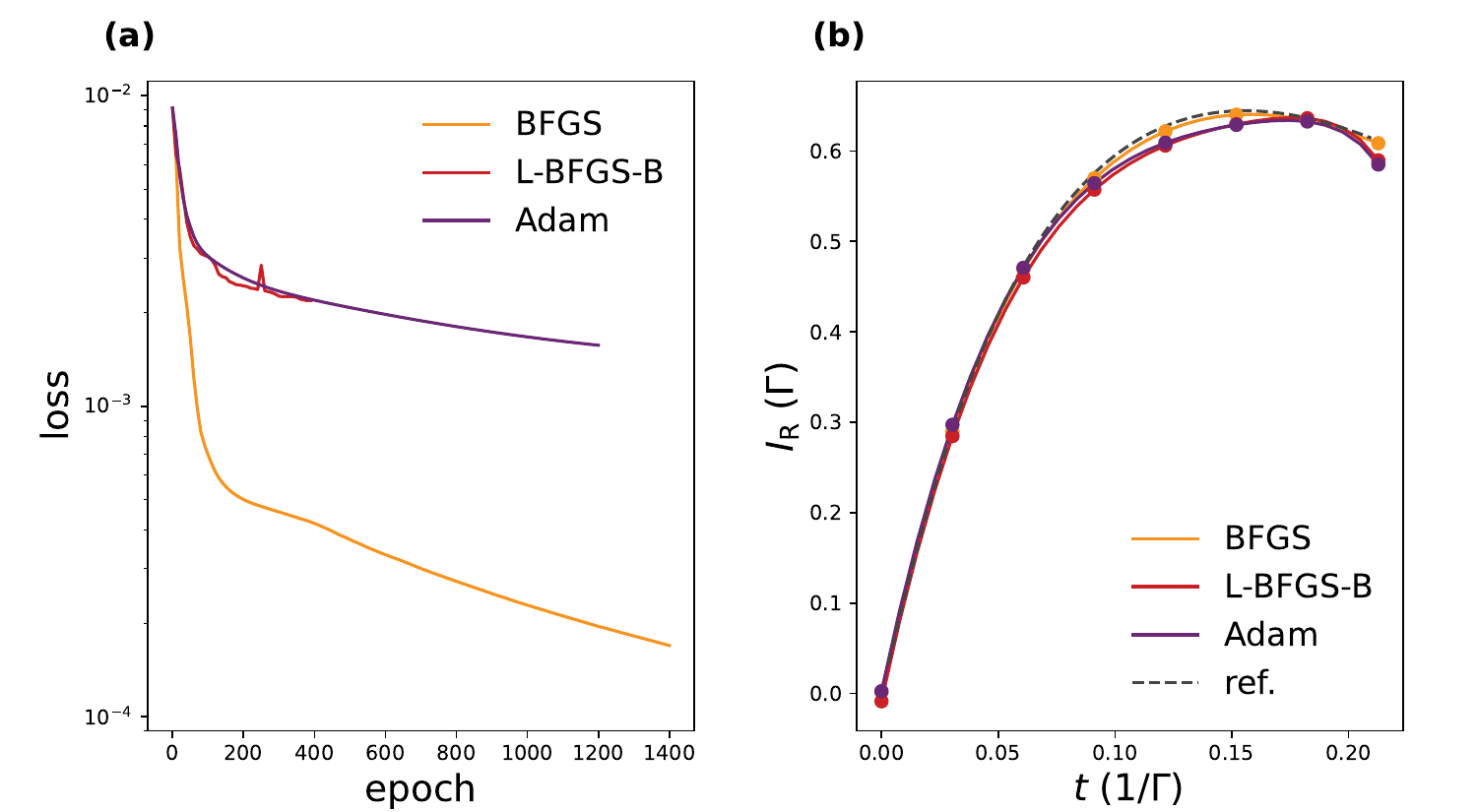}
	\caption{
    Comparison of different optimizers at the temperature $k_{\rm B}T=2.0\, \Gamma$ in the first subdomain $[0, 0.228\, \Gamma^{-1}]$.
    (a) Training loss (log scale) as a function of optimization steps.
    (b) Electric current obtained by different optimizers at the final points of (a).
    The mapping functions are $f_1(t)=t$, $f_2(t)=t^2$, and $f_3(t)=t^3$ for all cases. Dots on each curve mark the residual points used during optimization of PINN.
 } \label{fig:comparisonOp}
\end{figure}

Figure~\ref{fig:comparisonOp} compares the results of two common optimizers in PINN (Adam \cite{kingma2014adam} and L-BFGS-B \cite{zhu1997algorithm}) with that of the BFGS method. Figure~\ref{fig:comparisonOp}(a) shows the decreasing process of the training loss of different optimizers, illustrating that while BFGS reduces the loss to $10^{-4}$, the other two methods only achieve a loss of $10^{-3}$.  The L‑BFGS‑B method terminates early because the relative loss reduction fell below the convergence threshold. In contrast, the Adam method does not employ such a convergence criterion; instead, it continues training with a fixed learning rate of $4\times10^{-7}$. Figure~\ref{fig:comparisonOp}(b) compares the final results obtained by three optimizers. 
The results from the L-BFGS-B and Adam methods deviate significantly from the reference values, which will lead to large errors in subsequent evolution. Only the BFGS method yeiled the results that agree with the reference values, further confirming that the BFGS method is required for achieving sufficient accuracy.


\begin{thebibliography}{10}

\bibitem{Engel2007}
G.~S. Engel, T.~R. Calhoun, E.~L. Read, T.~Ahn, T.~Man{\v{c}}al, Y.~Cheng, R.~E. Blankenship, and G.~R. Fleming, \newblock ``Evidence for wavelike energy transfer through quantum coherence in photosynthetic systems,'' Nature {\bf 446}, 782 (2007).

\bibitem{doi:10.1021/jz900062f}
G.~D. Scholes, \newblock ``Quantum-Coherent Electronic Energy Transfer: Did Nature Think of It First?,'' J. Phys. Chem. Lett. {\bf 1}, 2 (2010).

\bibitem{doi:10.1126/science.1235820}
R.~Hildner, D.~Brinks, J.~B. Nieder, R.~J. Cogdell, and N.~F. van Hulst, \newblock ``Quantum Coherent Energy Transfer over Varying Pathways in Single Light-Harvesting Complexes,'' Science {\bf 340}, 1448 (2013).

\bibitem{6395f2ce-ffb5-3279-8695-df26ad69dc2d}
M.~Kasha, \newblock ``Energy Transfer Mechanisms and the Molecular Exciton Model for Molecular Aggregates,'' Radiat. Res. {\bf 20}, 55 (1963).

\bibitem{doi:10.1021/acs.chemrev.7b00581}
N.~J. Hestand and F.~C. Spano, \newblock ``Expanded Theory of H- and J-Molecular Aggregates: The Effects of Vibronic Coupling and Intermolecular Charge Transfer,'' Chem. Rev. {\bf 118}, 7069 (2018).

\bibitem{10.1063/1.5003306}
M.~Thoss and F.~Evers, \newblock ``{Perspective: Theory of quantum transport in molecular junctions},'' J. Chem. Phys. {\bf 148}, 030901 (2018).

\bibitem{Uzma2021}
F.~Uzma, L.~Yang, D.~He, X.~Wang, S.~Hu, L.~Ye, X.~Zheng, and Y.~Yan, \newblock ``Understanding the Sub-meV Precision-Tuning of Magnetic Anisotropy of Single-Molecule Junction,'' J. Phys. Chem. C {\bf 125}, 6990 (2021).

\bibitem{10.1063/1.4994987}
P.~Grégoire, A.~R. Srimath~Kandada, E.~Vella, C.~Tao, R.~Leonelli, and C.~Silva, \newblock ``{Incoherent population mixing contributions to phase-modulation two-dimensional coherent excitation spectra},'' J. Chem. Phys. {\bf 147}, 114201 (2017).

\bibitem{Bruder2018}
L.~Bruder, U.~Bangert, M.~Binz, D.~Uhl, R.~Vexiau, N.~Bouloufa-Maafa, O.~Dulieu, and F.~Stienkemeier, \newblock ``Coherent multidimensional spectroscopy of dilute gas-phase nanosystems,'' Nat. Commun. {\bf 9}, 4823 (2018).

\bibitem{Monroe2002}
C.~Monroe, \newblock ``Quantum information processing with atoms and photons,'' Nature {\bf 416}, 238 (2002).

\bibitem{head2020quantum}
K.~Head-Marsden, J.~Flick, C.~J. Ciccarino, and P.~Narang, \newblock ``Quantum information and algorithms for correlated quantum matter,'' Chem. Rev. {\bf 121}, 3061 (2020).

\bibitem{RevModPhys.94.045008}
I.~Piquero-Zulaica, J.~Lobo-Checa, Z.~M.~A. El-Fattah, J.~E. Ortega, F.~Klappenberger, W.~Auw\"arter, and J.~V. Barth, \newblock ``Engineering quantum states and electronic landscapes through surface molecular nanoarchitectures,'' Rev. Mod. Phys. {\bf 94}, 045008 (2022).

\bibitem{wilson1975renormalization}
K.~G. Wilson, \newblock ``The renormalization group: Critical phenomena and the Kondo problem,'' Rev. Mod. Phys. {\bf 47}, 773 (1975).

\bibitem{doi:10.1126/science.280.5363.567}
V.~Madhavan, W.~Chen, T.~Jamneala, M.~F. Crommie, and N.~S. Wingreen, \newblock ``Tunneling into a Single Magnetic Atom: Spectroscopic Evidence of the Kondo Resonance,'' Science {\bf 280}, 567 (1998).

\bibitem{doi:10.1126/science.1113449}
A.~Zhao, Q.~Li, L.~Chen, H.~Xiang, W.~Wang, S.~Pan, B.~Wang, X.~Xiao, J.~Yang, J.~G. Hou, and Q.~Zhu, \newblock ``Controlling the Kondo Effect of an Adsorbed Magnetic Ion Through Its Chemical Bonding,'' Science {\bf 309}, 1542 (2005).

\bibitem{li2020molecular}
X.~Li et~al., \newblock ``Molecular molds for regularizing Kondo states at atom/metal interfaces,'' Nat. Commun. {\bf 11}, 2566 (2020).

\bibitem{PhysRevLett.111.086601}
X.~Zheng, Y.~Yan, and M.~Di~Ventra, \newblock ``Kondo Memory in Driven Strongly Correlated Quantum Dots,'' Phys. Rev. Lett. {\bf 111}, 086601 (2013).

\bibitem{doi:10.1126/science.aay6779}
K.~Yang, W.~Paul, S.~H. Phark, P.~Willke, Y.~Bae, T.~Choi, T.~Esat, A.~Ardavan, A.~J. Heinrich, and C.~P. Lutz, \newblock ``Coherent spin manipulation of individual atoms on a surface,'' Science {\bf 366}, 509 (2019).

\bibitem{doi:10.1126/science.abg8223}
L.~M. Veldman, L.~Farinacci, R.~Rejali, R.~Broekhoven, J.~Gobeil, D.~Coffey, M.~Ternes, and A.~F. Otte, \newblock ``Free coherent evolution of a coupled atomic spin system initialized by electron scattering,'' Science {\bf 372}, 964 (2021).

\bibitem{Ding25084114}
X.~Ding, J.~Cao, X.~Zheng, and L.~Ye, \newblock ``Tracking spin flip-flop dynamics of surface molecules with quantum dissipation theory,'' J. Chem. Phys. {\bf 162}, 084114 (2025).

\bibitem{Tam18030402}
D.~Tamascelli, A.~Smirne, S.~F. Huelga, and M.~B. Plenio, \newblock ``Nonperturbative treatment of non-Markovian dynamics of open quantum systems,'' Phys. Rev. Lett. {\bf 120}, 030402 (2018).

\bibitem{Lam193721}
N.~Lambert, S.~Ahmed, M.~Cirio, and F.~Nori, \newblock ``Modelling the ultra-strongly coupled spin-boson model with unphysical modes,'' Nat. Commun. {\bf 10}, 3721 (2019).

\bibitem{PhysRevLett.82.1801}
W.~T. Strunz, L.~Di\'osi, and N.~Gisin, \newblock ``Open System Dynamics with Non-Markovian Quantum Trajectories,'' Phys. Rev. Lett. {\bf 82}, 1801 (1999).

\bibitem{10.1063/1.1647528}
J.~Shao, \newblock ``{Decoupling quantum dissipation interaction via stochastic fields},'' J. Chem. Phys. {\bf 120}, 5053 (2004).

\bibitem{PhysRevLett.123.050601}
L.~Han, V.~Chernyak, Y.~Yan, X.~Zheng, and Y.~Yan, \newblock ``Stochastic Representation of Non-Markovian Fermionic Quantum Dissipation,'' Phys. Rev. Lett. {\bf 123}, 050601 (2019).

\bibitem{PhysRevLett.88.170407}
J.~T. Stockburger and H.~Grabert, \newblock ``Exact $\mathit{c}$-Number Representation of Non-Markovian Quantum Dissipation,'' Phys. Rev. Lett. {\bf 88}, 170407 (2002).

\bibitem{suess2014hierarchy}
D.~Suess, A.~Eisfeld, and W.~T. Strunz, \newblock ``Hierarchy of stochastic pure states for open quantum system dynamics,'' Phys. Rev. Lett. {\bf 113}, 150403 (2014).

\bibitem{PhysRevLett.115.266802}
G.~Cohen, E.~Gull, D.~R. Reichman, and A.~J. Millis, \newblock ``Taming the Dynamical Sign Problem in Real-Time Evolution of Quantum Many-Body Problems,'' Phys. Rev. Lett. {\bf 115}, 266802 (2015).

\bibitem{PhysRevLett.130.186301}
A.~Erpenbeck, E.~Gull, and G.~Cohen, \newblock ``Quantum Monte Carlo Method in the Steady State,'' Phys. Rev. Lett. {\bf 130}, 186301 (2023).

\bibitem{PhysRevB.87.115115}
F.~G\"uttge, F.~B. Anders, U.~Schollw\"ock, E.~Eidelstein, and A.~Schiller, \newblock ``Hybrid NRG-DMRG approach to real-time dynamics of quantum impurity systems,'' Phys. Rev. B {\bf 87}, 115115 (2013).

\bibitem{PhysRevLett.88.256403}
M.~A. Cazalilla and J.~B. Marston, \newblock ``Time-Dependent Density-Matrix Renormalization Group: A Systematic Method for the Study of Quantum Many-Body Out-of-Equilibrium Systems,'' Phys. Rev. Lett. {\bf 88}, 256403 (2002).

\bibitem{RevModPhys.92.011001}
A.~U.~J. Lode, C.~L\'ev\^eque, L.~B. Madsen, A.~I. Streltsov, and O.~E. Alon, \newblock ``Colloquium: Multiconfigurational time-dependent Hartree approaches for indistinguishable particles,'' Rev. Mod. Phys. {\bf 92}, 011001 (2020).

\bibitem{10.1063/1.3173823}
H.~Wang and M.~Thoss, \newblock ``Numerically exact quantum dynamics for indistinguishable particles: The multilayer multiconfiguration time-dependent Hartree theory in second quantization representation,'' J. Chem. Phys. {\bf 131}, 024114 (2009).

\bibitem{RevModPhys.93.015008}
H.~Weimer, A.~Kshetrimayum, and R.~Or\'us, \newblock ``Simulation methods for open quantum many-body systems,'' Rev. Mod. Phys. {\bf 93}, 015008 (2021).

\bibitem{li2023dissipatons}
X.~Li, Y.~Su, Z.-H. Chen, Y.~Wang, R.-X. Xu, X.~Zheng, and Y.~Yan, \newblock ``Dissipatons as generalized Brownian particles for open quantum systems: Dissipaton-embedded quantum master equation,'' J. Chem. Phys. {\bf 158} (2023).

\bibitem{Yan14054105}
Y.~Yan, \newblock ``Theory of open quantum systems with bath of electrons and phonons and spins: Many-dissipaton density matrixes approach,'' J. Chem. Phys. {\bf 140}, 054105 (2014).

\bibitem{li2024toward}
X.~Li, S.~Lyu, Y.~Wang, R.~Xu, X.~Zheng, and Y.~Yan, \newblock ``Toward quantum simulation of non-Markovian open quantum dynamics: A universal and compact theory,'' Phys. Rev. A {\bf 110}, 032620 (2024).

\bibitem{moix2013hybrid}
J.~M. Moix and J.~Cao, \newblock ``A hybrid stochastic hierarchy equations of motion approach to treat the low temperature dynamics of non-Markovian open quantum systems,'' J. Chem. Phys. {\bf 139}, 134106 (2013).

\bibitem{duan2017zero}
C.~Duan, Z.~Tang, J.~Cao, and J.~Wu, \newblock ``Zero-temperature localization in a sub-ohmic spin-boson model investigated by an extended hierarchy equation of motion,'' Phys. Rev. B {\bf 95}, 214308 (2017).

\bibitem{hsieh2018unified}
C.-Y. Hsieh and J.~Cao, \newblock ``A unified stochastic formulation of dissipative quantum dynamics. I. Generalized hierarchical equations,'' J. Chem. Phys. {\bf 148}, 014103 (2018).

\bibitem{10.1063/1.1580111}
H.~Wang and M.~Thoss, \newblock ``{Multilayer formulation of the multiconfiguration time-dependent Hartree theory},'' J. Chem. Phys. {\bf 119}, 1289 (2003).

\bibitem{Tan89101}
Y.~Tanimura and R.~Kubo, \newblock ``Time evolution of a quantum system in contact with a nearly Gaussian-Markovian noise bath,'' J. Phys. Soc. Jpn. {\bf 58}, 101 (1989).

\bibitem{YAN2004216}
Y.~an~Yan, F.~Yang, Y.~Liu, and J.~Shao, \newblock ``Hierarchical approach based on stochastic decoupling to dissipative systems,'' Chem. Phys. Lett. {\bf 395}, 216 (2004).

\bibitem{Jin08234703}
J.~Jin, X.~Zheng, and Y.~Yan, \newblock ``{Exact dynamics of dissipative electronic systems and quantum transport: Hierarchical equations of motion approach},'' J. Chem. Phys. {\bf 128}, 234703 (2008).

\bibitem{PhysRevLett.109.266403}
Z.~Li, N.~Tong, X.~Zheng, D.~Hou, J.~Wei, J.~Hu, and Y.~Yan, \newblock ``Hierarchical Liouville-Space Approach for Accurate and Universal Characterization of Quantum Impurity Systems,'' Phys. Rev. Lett. {\bf 109}, 266403 (2012).

\bibitem{PhysRevA.55.2290}
B.~M. Garraway, \newblock ``Nonperturbative decay of an atomic system in a cavity,'' Phys. Rev. A {\bf 55}, 2290 (1997).

\bibitem{cirio2023pseudofermion}
M.~Cirio, N.~Lambert, P.~Liang, P.~Kuo, Y.~Chen, P.~Menczel, K.~Funo, and F.~Nori, \newblock ``Pseudofermion method for the exact description of fermionic environments: From single-molecule electronics to the Kondo resonance,'' Phys. Rev. Res. {\bf 5}, 033011 (2023).

\bibitem{Lin251289}
J.-D. Lin, P.-C. Kuo, N.~Lambert, A.~Miranowicz, F.~Nori, and Y.-N. Chen, \newblock ``Non-Markovian quantum exceptional points,'' Nat. Commun. {\bf 16}, 1289 (2025).

\bibitem{PhysRevResearch.2.043058}
G.~Pleasance, B.~M. Garraway, and F.~Petruccione, \newblock ``Generalized theory of pseudomodes for exact descriptions of non-Markovian quantum processes,'' Phys. Rev. Res. {\bf 2}, 043058 (2020).

\bibitem{verstraete2008matrix}
F.~Verstraete, V.~Murg, and J.~I. Cirac, \newblock ``Matrix product states, projected entangled pair states, and variational renormalization group methods for quantum spin systems,'' Advances in physics {\bf 57}, 143 (2008).

\bibitem{schollwock2011density}
U.~Schollw{\"o}ck, \newblock ``The density-matrix renormalization group in the age of matrix product states,'' Annals of physics {\bf 326}, 96 (2011).

\bibitem{orus2014practical}
R.~Or{\'u}s, \newblock ``A practical introduction to tensor networks: Matrix product states and projected entangled pair states,'' Annals of physics {\bf 349}, 117 (2014).

\bibitem{shi2018efficient}
Q.~Shi, Y.~Xu, Y.~Yan, and M.~Xu, \newblock ``Efficient propagation of the hierarchical equations of motion using the matrix product state method,'' J. Chem. Phys. {\bf 148}, 174102 (2018).

\bibitem{doi:10.1126/science.aag2302}
G.~Carleo and M.~Troyer, \newblock ``Solving the quantum many-body problem with artificial neural networks,'' Science {\bf 355}, 602 (2017).

\bibitem{PhysRevLett.125.100503}
M.~Schmitt and M.~Heyl, \newblock ``Quantum Many-Body Dynamics in Two Dimensions with Artificial Neural Networks,'' Phys. Rev. Lett. {\bf 125}, 100503 (2020).

\bibitem{PhysRevLett.122.250502}
M.~J. Hartmann and G.~Carleo, \newblock ``Neural-Network Approach to Dissipative Quantum Many-Body Dynamics,'' Phys. Rev. Lett. {\bf 122}, 250502 (2019).

\bibitem{PhysRevLett.122.250501}
A.~Nagy and V.~Savona, \newblock ``Variational Quantum Monte Carlo Method with a Neural-Network Ansatz for Open Quantum Systems,'' Phys. Rev. Lett. {\bf 122}, 250501 (2019).

\bibitem{PhysRevLett.122.250503}
F.~Vicentini, A.~Biella, N.~Regnault, and C.~Ciuti, \newblock ``Variational Neural-Network Ansatz for Steady States in Open Quantum Systems,'' Phys. Rev. Lett. {\bf 122}, 250503 (2019).

\bibitem{PhysRevB.99.214306}
N.~Yoshioka and R.~Hamazaki, \newblock ``Constructing neural stationary states for open quantum many-body systems,'' Phys. Rev. B {\bf 99}, 214306 (2019).

\bibitem{cao2026simulating}
L.~Cao, L.~Ge, D.~Zhang, X.~Li, J.~Pan, Y.~Wang, R.-X. Xu, Y.~Yan, and X.~Zheng, \newblock ``Simulating non-Markovian open quantum dynamics with neural quantum states,'' Physical Review B {\bf 113}, L140301 (2026).

\bibitem{ye2025simulating}
L.~Ye, Y.~Wang, and X.~Zheng, \newblock ``Simulating many-body open quantum systems by harnessing the power of artificial intelligence and quantum computing,'' J. Chem. Phys. {\bf 162}, 120901 (2025).

\bibitem{ke2022hierarchical}
Y.~Ke, R.~Borrelli, and M.~Thoss, \newblock ``Hierarchical equations of motion approach to hybrid fermionic and bosonic environments: Matrix product state formulation in twin space,'' J. Chem. Phys. {\bf 156}, 194102 (2022).

\bibitem{ke2023tree}
Y.~Ke, \newblock ``Tree tensor network state approach for solving hierarchical equations of motion,'' J. Chem. Phys. {\bf 158}, 211102 (2023).

\bibitem{preston2025nonadiabatic}
R.~J. Preston, Y.~Ke, S.~L. Rudge, N.~Hertl, R.~Borrelli, R.~J. Maurer, and M.~Thoss, \newblock ``Nonadiabatic quantum dynamics of molecules scattering from metal surfaces,'' J. Chem. Theory Comput. {\bf 21}, 1054 (2025).

\bibitem{zhang2024artificial}
J.~Zhang, C.~L. Benavides-Riveros, and L.~Chen, \newblock ``Artificial-intelligence-based surrogate solution of dissipative quantum dynamics: physics-informed reconstruction of the universal propagator,'' The Journal of Physical Chemistry Letters {\bf 15}, 3603 (2024).

\bibitem{zhang2025neural}
J.~Zhang, C.~L. Benavides-Riveros, and L.~Chen, \newblock ``Neural network solution of non-Markovian quantum state diffusion and operator construction of quantum stochastic process,'' The Journal of Chemical Physics {\bf 163} (2025).

\bibitem{reh2021time}
M.~Reh, M.~Schmitt, and M.~G{\"a}rttner, \newblock ``Time-dependent variational principle for open quantum systems with artificial neural networks,'' Phys. Rev. Lett. {\bf 127}, 230501 (2021).

\bibitem{raissi2018deep}
M.~Raissi, \newblock ``Deep hidden physics models: Deep learning of nonlinear partial differential equations,'' J. Mach. Learn. Res. {\bf 19}, 1 (2018).

\bibitem{liu2021dual}
D.~Liu and Y.~Wang, \newblock ``A dual-dimer method for training physics-constrained neural networks with minimax architecture,'' Neural Networks {\bf 136}, 112 (2021).

\bibitem{kharazmi2021hp}
E.~Kharazmi, Z.~Zhang, and G.~E. Karniadakis, \newblock ``hp-VPINNs: Variational physics-informed neural networks with domain decomposition,'' Computer Methods in Applied Mechanics and Engineering {\bf 374}, 113547 (2021).

\bibitem{jagtap2020conservative}
A.~D. Jagtap, E.~Kharazmi, and G.~E. Karniadakis, \newblock ``Conservative physics-informed neural networks on discrete domains for conservation laws: Applications to forward and inverse problems,'' Computer Methods in Applied Mechanics and Engineering {\bf 365}, 113028 (2020).

\bibitem{cuomo2022scientific}
S.~Cuomo, V.~S. Di~Cola, F.~Giampaolo, G.~Rozza, M.~Raissi, and F.~Piccialli, \newblock ``Scientific machine learning through physics--informed neural networks: Where we are and what’s next,'' J. Sci. Comput. {\bf 92}, 88 (2022).

\bibitem{sinibaldi2026time}
A.~Sinibaldi, D.~Hendry, F.~Vicentini, and G.~Carleo, \newblock ``Time-dependent neural Galerkin method for quantum dynamics,'' Physical Review Letters {\bf 136}, 120402 (2026).

\bibitem{van2025many}
A.~Van~de Walle, M.~Schmitt, and A.~Bohrdt, \newblock ``Many-body dynamics with explicitly time-dependent neural quantum states,'' Machine Learning: Science and Technology {\bf 6}, 045011 (2025).

\bibitem{ullah2024physics}
A.~Ullah, Y.~Huang, M.~Yang, and P.~O. Dral, \newblock ``Physics-informed neural networks and beyond: enforcing physical constraints in quantum dissipative dynamics,'' Digital Discovery {\bf 3}, 2052 (2024).

\bibitem{ullah2025machine}
A.~Ullah and J.~O. Richardson, \newblock ``Machine learning meets su (n) Lie algebra: Enhancing quantum dynamics learning with exact trace conservation,'' The Journal of Chemical Physics {\bf 162} (2025).

\bibitem{atif2026toward}
M.~Atif, A.~Ullah, and M.~Yang, \newblock ``Toward quantum-aware machine learning: Improved prediction of quantum dissipative dynamics via complex valued neural networks,'' The Journal of Chemical Physics {\bf 164} (2026).

\bibitem{dugan2023q}
O.~M. Dugan, P.~Y. Lu, R.~Dangovski, D.~Luo, and M.~Soljacic,
\newblock Q-flow: generative modeling for differential equations of open quantum dynamics with normalizing flows,
\newblock in {\em International Conference on Machine Learning}, pages 8879--8901, PMLR, 2023.

\bibitem{And6141}
P.~W. Anderson, \newblock ``Localized Magnetic States in Metals,'' Phys. Rev. {\bf 124}, 41 (1961).

\bibitem{hu2010communication}
J.~Hu, R.-X. Xu, and Y.~Yan, \newblock ``Communication: Pad{\'e} spectrum decomposition of Fermi function and Bose function,'' J. Chem. Phys. {\bf 133}, 101106 (2010).

\bibitem{Hou15104112}
D.~Hou, S.~K. Wang, R.~L. Wang, L.~Z. Ye, R.~X. Xu, X.~Zheng, and Y.~J. Yan, \newblock ``Improving the efficiency of hierarchical equations of motion approach and application to coherent dynamics in Aharonov-Bohm interferometers,'' J. Chem. Phys. {\bf 142}, 104112 (2015).

\bibitem{krishnapriyan2021characterizing}
A.~Krishnapriyan, A.~Gholami, S.~Zhe, R.~Kirby, and M.~W. Mahoney, \newblock ``Characterizing possible failure modes in physics-informed neural networks,'' Adv. Neural Inf. Process. Syst. {\bf 34}, 26548 (2021).

\bibitem{robbins1951stochastic}
H.~Robbins and S.~Monro, \newblock ``A stochastic approximation method,'' Ann. Math. Stat. {\bf 22}, 400 (1951).

\bibitem{kingma2014adam}
D.~P. Kingma and J.~Ba, \newblock ``Adam: A method for stochastic optimization,'' arXiv preprint arXiv:1412.6980  (2014).

\bibitem{loshchilov2017decoupled}
I.~Loshchilov and F.~Hutter, \newblock ``Decoupled weight decay regularization,'' arXiv preprint arXiv:1711.05101  (2017).

\bibitem{fletcher2013practical}
R.~Fletcher,
\newblock {\em Practical Methods of Optimization},
\newblock John Wiley \& Sons, 2013.

\bibitem{ye2016heom}
L.~Ye, X.~Wang, D.~Hou, R.-X. Xu, X.~Zheng, and Y.~Yan, \newblock ``HEOM-QUICK: a program for accurate, efficient, and universal characterization of strongly correlated quantum impurity systems,'' Wiley Interdisciplinary Reviews: Computational Molecular Science {\bf 6}, 608 (2016).

\bibitem{zhang2024heom}
D.~Zhang, L.~Ye, J.~Cao, Y.~Wang, R.-X. Xu, X.~Zheng, and Y.~Yan, \newblock ``HEOM-QUICK2: A general-purpose simulator for fermionic many-body open quantum systems --- An update,'' Wiley Interdisciplinary Reviews: Computational Molecular Science {\bf 14}, e1727 (2024).

\bibitem{li2024quantum}
X.~Li, S.-X. Lyu, Y.~Wang, R.-X. Xu, X.~Zheng, and Y.~Yan, \newblock ``Toward quantum simulation of non-Markovian open quantum dynamics: A universal and compact theory,'' Phys. Rev. A {\bf 110}, 032620 (2024).

\bibitem{he2016deep}
K.~He, X.~Zhang, S.~Ren, and J.~Sun,
\newblock Deep residual learning for image recognition,
\newblock in {\em Proceedings of the IEEE conference on computer vision and pattern recognition}, pages 770--778, 2016.

\bibitem{cheng2021deep}
C.~Cheng and G.-T. Zhang, \newblock ``Deep learning method based on physics informed neural network with resnet block for solving fluid flow problems,'' Water {\bf 13}, 423 (2021).

\bibitem{Schuld19}
M.~Schuld, I.~Sinayskiy, and F.~Petruccione, \newblock ``Neural Networks Take on Open Quantum Systems,'' Physics {\bf 12}, 74 (2019).

\bibitem{zhu1997algorithm}
C.~Zhu, R.~H. Byrd, P.~Lu, and J.~Nocedal, \newblock ``Algorithm 778: L-BFGS-B: Fortran subroutines for large-scale bound-constrained optimization,'' ACM Transactions on mathematical software (TOMS) {\bf 23}, 550 (1997).

\end{thebibliography}

\end{document}